# Operando characterization of interfacial charge transfer processes

Christoph Baeumer[1,2]

1 MESA+ Institute for Nanotechnology, University of Twente, Faculty of Science and Technology, P.O. Box 217, 7500 AE Enschede, Netherland

2 Peter Gruenberg Institute and JARA-FIT, Forschungszentrum Juelich GmbH, 52425 Juelich, Germany

## Abstract

Interface science has become a key aspect for fundamental research questions and for the understanding, design and optimization of urgently needed energy and information technologies. As the interface properties change during operation, e.g. under applied electrochemical stimulus, and because multiple bulk and interface processes coexist and compete, detailed operando characterization is needed. In this perspective, I present an overview of the state-of-the art and challenges in selected X-ray spectroscopic techniques, concluding that among others, interface-sensitivity remains a major concern in the available techniques. I propose and discuss a new method to extract interface-information from nominally bulk sensitive techniques, and critically evaluate the selection of X-ray energies for the recently developed meniscus X-ray photoelectron spectroscopy, a promising operando tool to characterize the solid-liquid interface. I expect that these advancements along with further developments in time and spatial resolution will expand our ability to probe the interface electronic and molecular structure with sub-nm depth and complete our understanding of charge transfer processes during operation.

# 1 Introduction and background

Interfaces are the cornerstone in a plethora of current and emerging technologies, and in fundamental future research directions. They are a platform for exploiting extraordinary phenomena that result from reduced dimensions or proximity of dissimilar materials. At the same time, interfaces can present the bottleneck in technologies relying on charge transfer processes. These challenges and opportunities require experimental probes to shed light on the underlying physical and chemical processes.

One can differentiate the types of interfaces based on the state of matter and the types of charge transfer processes. Here, we focus on the solid-solid, solid-gas and solid-liquid interfaces and distinguish between electrostatic and electrochemical charge transfer. Particular attention will be given to the characterization of electrochemical solid-liquid interfaces, because of their urgent relevance to address societal challenges. Today, electrochemical solid-liquid interfaces govern applications in, inter alia, sensing, chemical manufacturing, and, most urgently, energy conversion and storage. The central role of the electrochemical solid-liquid interface has already been identified in the 1800's.[1,2] The description of electronic and molecular structure of such interfaces was first attempted in the 19$^{th}$ and 20$^{th}$ centuries,[3] and it is still under refinement.[4–6]

The continuing pursuit of a fundamental understanding of the molecular-level structure and dynamic processes like electronic and ionic charge accumulation and transfer across the interface is complicated by inherent experimental challenges in the interface-sensitive characterization of chemical and electronic states.[7] Our most complete understanding of the solid-liquid interface has been derived from the investigation of the liquid molecular[6,8] and electronic[5] structure near comparably simple noble metal electrodes. More complex but industrially more attractive solids like transition metal oxides and carbides, however, are much harder to understand at the atomic-scale: they exhibit an intricate set of electrochemical phenomena including bulk ion intercalation alongside several coexisting reactions at the interface, which need to be separated by experimental probes under operating conditions.

Here, I will first introduce examples of interfacial charge transfer processes of interest, followed by a brief and non-exhaustive summary of available experimental approaches for the operando interface characterization. I will focus on X-ray spectroscopic techniques with special attention on recent developments for operando X-ray photoelectron spectroscopy of the solid-liquid interface and briefly mention optical and vibrational spectroscopies. It will become clear that interface-sensitivity is the crucial concern in most techniques. This will form the basis for my perspective on future developments and experimental avenues for interface-sensitive X-ray spectroscopic techniques to obtain a similar level of understanding of the solid-liquid interface as has been achieved for the solid-solid and solid-gas interfaces.

## 2 Selected electronic and ionic interfacial charge transfer processes

Charge transfer processes are ubiquitous in interface science. This perspective will focus on the transfer of electronic or ionic charge carriers across the interface between two materials and summarize a few notable and instructional example phenomena. Purely electronic charge transfer can be viewed as exchange of electrons (or holes) across the interface driven by the alignment of the Fermi level (the electrochemical potential of electrons). Ionic charge transfer refers to ions (or ionic defects) crossing the interface driven by the alignment of their electrochemical potential.[9] Generally, we aim at understanding and controlling ionic and electronic structures at various interfaces to unlock the next generation of electronic and electrochemical devices for energy conversion and storage, sensing, and brain inspired computing, to name a few.

### 2.1 Solid-solid interfaces

For the solid-solid interface, the electronic charge-transfer across oxide interfaces has become a major platform to leverage nanoscale phenomena and induce new properties, e.g. to stimulate electronic conduction in nominally insulating materials or magnetism in non-magnetic materials.[10–12] An elegant "remote doping" scheme was theoretically predicted[13,14] and experimentally verified[12,14,15] for perovskite oxide $ABO_3$/$AB`O_3$ heterointerfaces with transition metal B sites. For example, $LaTiO_3$/$LaCoO_3$ exhibits a valence change from $Co^{3+}$ to $Co^{2+}$ resulting from electron transfer based on the alignment of O 2p-related bands in a continuous backbone of transition metal oxide octahedra across the interface.[15] The electronic charge transfer across a perovskite oxide interface is schematically depicted in Figure 1a.

Another, widely studied charge transfer phenomena is the so-called polar catastrophe at interfaces between polar and non-polar perovskite oxides, e.g. between $LaAlO_3$ and $SrTiO_3$, which may be resolved by electronic charge transfer, resulting in a valence change from $Ti^{4+}$ to $Ti^{3+}$.[10,11] In addition to the electron transfer ionic charge transfer across the interface must also be taken into account. For example, cation vacancies,[16,17] oxygen vacancies[18] and ionic intermixing[19] have been observed at the $LaAlO_3$/$SrTiO_3$ interface.

For ionic charge transfer across solid-solid interfaces, all solid state batteries present one of the major recent research thrusts.[20] They promise improvements in battery safety and lifetime, as well as higher energy and power densities. Generally, there are numerous interfaces inside batteries, such as the electrode-electrolyte interface where ion intercalation occurs, and homogeneous interfaces between electrode particles. These interfaces are often the limiting factor in battery performance, because of slow ionic migration across interfaces and growth of unwanted interfacial layers.[20,21] More details on interfaces in solid-state batteries can be found in recent dedicated reviews.[21,22]

Another interesting example for ionic charge transfer across a solid-solid interface is the so-called resistive switching or memristive effect. Here, oxygen vacancies are created through local redox reactions at metal/oxide or oxide/oxide interfaces and oxygen ions migrate reversibly under an applied electric field,[23–25] either across the entire interface or within a confined region (called switching filament) at the interface. Because the electronic conductivity of metal/oxide interfaces in these devices depends strongly (typically exponentially) on the oxygen vacancy concentration, this phenomenon can be used as an analogue or digital switch, presenting one of the most attractive pathways for brain-inspired computing architectures. The oxygen vacancy migration at an oxide/oxide/metal interface is depicted schematically in Figure 1b. Details of the switching process have been reviewed in detail elsewhere.[26,27]

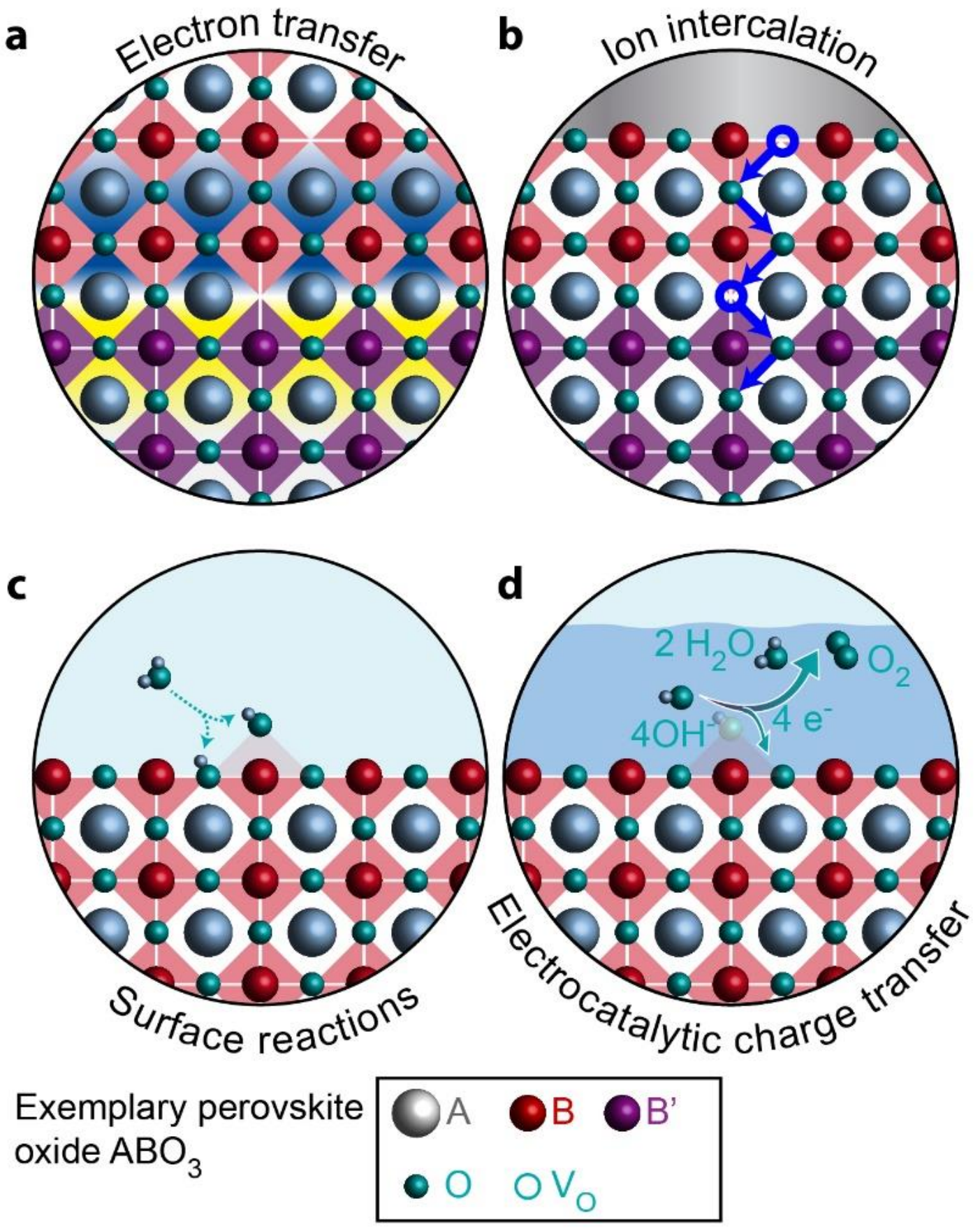

**Figure 1:** Schematic charge transfer processes for exemplary interfaces based on perovskite oxides. a) Electron transfer across an $ABO_3$/$AB'O_3$ heterointerface. Gain (loss) of electron density is shown in blue (yellow). b) Ionic transport across interfaces, for example during resistive switching in an oxide heterostructure with a metal top electrode (grey). Oxygen vacancies move across the solid-solid interface due to an external electric field. c) Hydroxylation of an oxide surface exposed to a water containing atmosphere. d) Schematic oxygen evolution reaction on an oxide surface in alkaline media.

## 2.2 Solid-gas interfaces

The solid-gas interface is particularly attractive for (electro-)catalytic reactions, e.g. in solid-oxide fuel cells,[28–30] sensing applications, and chemical synthesis such as the Fischer-Tropsch method[31] to produce petroleum products from nonpetroleum feedstock. For these cases, the adsorption, surface (redox) reaction and desorption are of particular importance. During these processes, charge can be transferred between the adsorbing gaseous species and the solid surface. Figure 1c shows the simplified example of hydroxylation of an oxide surface in the presence of water vapor. The solid surface exposed to different atmospheres may result in specific defects, even in highly ordered surfaces, e.g. for titanium dioxide [32] and perovskite oxides [33,34] and the insights about the resulting surface and defect structures can be used to optimize, e.g., the catalytic performance.[35]

## 2.3 Solid-liquid interfaces

The structure and resulting properties of the solid-liquid interface, are more complex and challenging to probe experimentally compared to the solid/solid and solid/gas interfaces. This is due to low-temperature surface transformations of the solid, which are typically accompanied by a loss of long-range order,[36,37] and due to challenges in characterization, as addressed in detail in sections 3 and 4. The charge transfer processes at the solid-liquid interface are similar to the solid-gas counterpart and are very diverse: adsorption and desorption of (ionic) species, surface redox reactions in the solid electrode or in the liquid electrolyte, electrocatalysis and ion intercalation, to name a few. They enable various key technologies, including lithium-ion batteries,[38] supercapacitors,[39] electrocatalysis,[40] photocatalysis,[23] and electroplating. Figure 1d schematically shows the electrocatalytic oxygen evolution reaction at a solid-liquid interface, i.e. the kinetically limiting half-cell reaction for hydrogen production through water electrolysis.

## 2.4 The need for operando characterization

To optimize all these applications, an advance in fundamental knowledge is prerequisite. But the interfacial processes are difficult to understand in detail, especially for the solid-liquid interface. Typically, several processes coexist in the same materials and under similar conditions, either competing with or assisting one another. Additionally, a complex electrostatic double layer forms at the interface when a solid is immersed in a liquid. Electrochemical processes like ion insertion or surface redox reactions precede electrocatalytic reactions[41] or occur at almost the same potential,[42] resulting in transformed surface and bulk properties under operating conditions.[37]

These charge transfer processes occurring at the interface and in the bulk and the resulting chemical and physical properties can be fully reversible: They cannot be probed ex situ because the interface may transform as soon as external stimuli are removed. Instead, they must be characterized under operating conditions (“operando”) to overcome the limits in our understanding and ultimately enable efficient utilization. Significant experimental and conceptual progress has been made in recent years, yet experimental challenges remain. E.g., the relevant or performance-limiting processes often occur at the interface itself or within nanometer-sized interfacial layers, resulting in small signals for many experimental probes, which are often overshadowed by signal from the bulk. The examples above also highlight that the complex interplay of different species at various interfaces necessitate characterization of multiple properties preferably simultaneously. For example, the nanoscale chemical composition determines the *electronic* structure in memristive devices. Lastly, understanding one of such interfaces may also unlock new properties at a second type of interfaces. For example, the groups of Koster and Golden suggested to use electronic charge transfer at solid-solid interfaces to tune the activity for electrocatalysis at the solid-liquid interfaces.[15] Therefore, we need interface-sensitive and interface-selective operando probes that collect interpretable signal from the interface of interest without overshadowing by the bulk solid or liquid.

# 3 Selected state-of-the-art operando characterization approaches

Many different techniques to probe various interfaces under operating conditions have matured over the past decades or years, each with its own advantages and shortcomings. Selected techniques are shown in Table 1. Throughout this perspective, I will focus on X-ray spectroscopic techniques and briefly mention optical and vibrational spectroscopies. Other, undoubtedly also important and promising techniques for the study of various interfacial processes are beyond the scope of this perspective. Several insightful reviews on the topic can be found in references [43–50].

Generally, photon and electron probes are attractive probes of interfaces, because photon-matter and electron-matter interactions happen much faster than dynamic Brownian motion, allowing characterization of "frozen snapshots" of the dynamic interfaces.[7] Without special efforts for pump-probe approaches, these techniques yield a steady-state, average depiction of the interface of interest. In this section, I will highlight a non-exhaustive selection of exemplary spectroscopic techniques that enabled in-depth understanding of interfacial charge transfer processes in recent years.

**Table 1: Selected operando characterization tools**

| Technique | Probe/detected species | Sensitive for | Advantages | Disadvantages | Further Reading |
|---|---|---|---|---|---|
| Vibrational and optical spectroscopic techniques | | | | | |
| **Infrared spectroscopy** | Photons in, photons out | Functional groups, specific bonds | Comparably simple experimental setup. | Not necessarily interface-sensitive | 51 |
| **Raman spectroscopy** | Photons in, photons out | Functional groups, specific bonds | Comparably simple experimental setup. Complementary to IR. | Not necessarily interface-sensitive | 52 |
| **Sum frequency generation spectroscopy** | Photons in, photons out | Functional groups, specific bonds | Interface-sensitive | Complex experimental design | 42,53,54 |
| **UV-Vis spectroscopy** | Photons in, photons out | Oxidation state, phase, composition | Comparably simple experimental setup | Typically not interface-sensitive | [55–57] |
| X-ray Spectroscopic techniques | | | | | |
| **Hard X-ray absorption spectroscopy** | Photons in, electrons or photons out | Atomic concentrations, oxidation states, local geometries | Comparably simple experimental cell | Not interface-sensitive enough. Usually requires synchrotron radiation | [7,38,40,58] |
| **Soft X-ray absorption spectroscopy** | Photons in, electrons or photons out | Atomic concentrations, oxidation states, local geometries | Very sensitive for oxidation state and local geometry | Not interface-sensitive enough. Usually requires synchrotron radiation and complicated experimental setups | [7,38,40,58,59] |
| **Hard X-ray photoelectron spectroscopy** | Photons in, electrons out | Atomic concentrations, oxidation states, and electrostatic potentials | Sub-surface sensitive (up to 10s of nm information depth) | Not interface-sensitive enough. Requires special X-ray sources or synchrotron radiation | [60–63] |
| **Soft X-ray photoelectron spectroscopy** | Photons in, electrons out | Atomic concentrations, oxidation states and electrostatic potentials | Surface sensitive (0.5-2 nm mean information depth) | Limited information depth. Need for UHV | [64–70] |
| **Near-ambient pressure X-ray photoelectron spectroscopy** | Photons in, electrons out | Atomic concentrations, oxidation states and electrostatic potentials as a function of temperature and pressure | Surface sensitive (0.5-2 nm mean information depth). Sensitive for band alignments | Limited information depth | [44,71–74] |
| **Meniscus X-ray photoelectron spectroscopy** | Photons in, electrons out | Atomic concentrations, oxidation states and double layer potential | Solid material of any thickness interfaced with a liquid. Sensitive for band alignments at the interface | Mass and charge transport limitations. Limited information depth or limitations in signal-to-noise ratio. Meniscus instability | [75–79] |
| **XAS and XPS with thin membranes** | Photons in, electrons or photons out | Local structure, atomic concentrations, oxidation states and electrostatic potentials | Avoid limitations from mass transport using a flow cell setup. Can be interface-sensitive | Limitation to selected materials and geometries. Risk of membrane failure | [43,48,80] |
| **Photoemission electron microscopy with membranes** | Photons in, electrons out | Atomic concentrations, oxidation states and electrostatic potentials with spatial resolution | Spatial resolution, "multiple samples simultaneously" | Difficult sample fabrication and risk of bursting | [81–84] |
| **Standing wave X-ray photoelectron spectroscopy and X-ray absorption spectroscopy** | Photons in, electrons or photons out | Atomic concentrations, oxidation states and double layer potential with extreme depth resolution | Highest depth resolution (Ångström-scale) | Complicated samples, long measurement times, complex analysis | [85–88] |
| Selection of complementary techniques not covered in this perspective | | | | | |
| **Scanning probe microscopy** | Scanning probes of various designs | Morphology, atomic surface structure, electrostatics, spatially resolved electrochemical activity | Versatile platform capable of high spatial resolution | No direct probe of chemical composition and oxidation states | [89–94] |
| **Mössbauer spectroscopy** | Gamma radiation | spin and/or oxidation states | Very sensitive for small changes | Limited to very few elements like Fe, I, Sn, and Sb | [95] |
| **X-ray emission spectroscopy** | Photons in, electrons out | Electronic structure | Complementary to XAS | Not interface-sensitive | [96,97] |
| **Surface X-ray diffraction** | Photons in, electrons out | Surface structure, adsorbed species, structure of the liquid layer | Very sensitive for the interface structure | Requires extensive modelling and prior knowledge about the surface structure. Usually requires synchrotron radiation. | [8,98–103] |
| **Resonant inelastic X-ray scattering** | Photons in, electrons out | occupied states, charge transfer, low-energy excitations, complementary to XAS | Two-dimensional data maps and the high resolution in the energy transfer. | Usually not interface-sensitive. Requires synchrotron radiation. | [49,59,96,104] |
| **Transmission electron microscopy** | Electrons in, electrons out | Structure, composition, oxidation states, electric fields | Highest spatial resolution. Capable of structural, electronic, magnetic and chemical mapping | Limited to thin (electron-transparent) samples. Difficult sample fabrication and high risk of fabrication-induced structural changes | [24,49,105–109] |
| **Electron Paramagnetic Resonance** | Microwaves | Unpaired electrons or radicals | Sensitive for reaction intermediates | Not necessarily interface-sensitive | [110,111] |
| **Nuclear magnetic resonance spectroscopy** | Radiowaves | Structure and composition | High sensitivity | Low signals | [112] |
| **Neutron reflectometry** | Neutrons in, neutrons out | Composition and structure of interfaces | Sensitive to light elements. Can probe buried interfaces. Possibility of isotopic labelling | Need for neutrons | [113] |

## 3.1 Optical and vibrational spectroscopy

Vibrational spectroscopies probe the vibrational energy of chemical bonds and specific functional groups and can yield detailed information about bulk structures and adsorbed species at interfaces. Prominent examples are the complementary techniques infrared spectroscopy and Raman spectroscopy. Infrared spectroscopy probes the absorption due to vibrational modes and is used to identify chemical substances or track functional groups from absorption, emission, or reflection of infrared light, while Raman spectroscopy measures the energy difference between an incident photon and the scattered photon after an inelastic scattering process. Summaries about both techniques are provided in [51] and [52].

As with other operando spectroscopy tools discussed below, the interface-sensitivity is an important aspect in vibrational spectroscopy, which is nominally bulk-sensitive. Another main limitation results from spectral interference from gas molecules in the experimental setup like water vapor. Dedicated measures to reduce such interference effects and enhance interface-sensitivity are required, for example through (polarization modulation) infrared reflection–adsorption approaches.[114,115] To achieve maximum interface-sensitivity, special sample geometries can be used to achieve so-called surface-enhanced infrared[116] and Raman spectra.[117,118] Examples include the identification of active oxygen sites resulting from the surface deprotonation process in $Ni(OH)_2$/NiOOH electrocatalysts,[56] catalyst-adsorbate interactions[119], structural changes in adsorbed species,[120] and charge transfer at core-shell nanoparticles.[118] In addition, sum frequency generation spectroscopy is a particularly successful vibrational spectroscopy tool for interface characterization because of the selection rules governing the underlying non-linear optical processes. [42,53,54]

UV-Vis spectroscopy[55] is based on the absorption of ultraviolet and visible light by molecules or solids due to low-energy electronic excitations from the ground state to excited states (typically from the valence band to the conduction band in solids) and can be performed in transmission or reflection mode. Generally, UV-Vis spectroscopy is a bulk-sensitive technique, and it has found wide-spread application to identify the oxidation state of organic and inorganic materials. These include dyes because UV-Vis spectroscopy essentially probes the perceived color of a given substance, and electrochemical materials like $Ni(OH)_2$/NiOOH-based electrocatalysts, where UV-Vis spectroscopy revealed the active phase under operating conditions.[56,57] Recently we accomplished extraction of interface-sensitive information from nominally bulk-sensitive UV-Vis spectra, as will be discussed in detail in section 4.2.[37]

## 3.2 X-ray absorption spectroscopy

In X-ray absorption spectroscopy (XAS), an electron is excited also from a ground state into an empty, excited state. It thus probes the unoccupied part of the electronic structure. The difference to UV-Vis spectroscopy lies in the involved electronic states: The energy difference in UV-Vis is in the range of few eV, while it is hundreds or thousands of eV for XAS, because electrons are excited from the core levels rather than from the valence band. So-called absorption edges are characterized by the energy difference between a specific core level and a specific unoccupied electronic state. The detailed formalism based on Fermi's golden rule is described, e.g., in the review by de Groot[97] and the book by Stöhr.[121] XAS is element-specific and yields information about the chemical surrounding of each component. Typically, one distinguishes between the near-edge region (up to 50 eV above the edge, X-ray absorption near edge spectroscopy (XANES)), and the extended structure (more than 50 eV above the edge, extended X-ray absorption fine structure (EXAFS)), which exhibits an oscillatory structure in the X-ray absorption coefficient. XANES is sensitive for the oxidation state, coordination geometry and number, and elemental composition, while EXAFS is mostly used for determination of local structural information such as the distance of neighboring atoms.

Traditionally, X-ray absorption spectra were determined directly from the transmission intensity of X-rays penetrating a thin specimen.[122] But indirect measurements relying on fluorescence yield or secondary or Auger electrons have become dominant[121] and synchrotron facilities with tunable and high-brilliance X-rays have enabled the development of operando X-ray absorption spectroscopy.[7] Generally, XAS is not an interface-sensitive technique, but the information depth depends critically on the detection mode: detection of transmitted X-rays or fluorescence can be considered bulk-sensitive with hundreds of nanometers information depth. For the detection of partial or total electron yield, the information depth can be in the range of 1-10 nm.[38] Another useful option is the use of grazing incidence (or grazing exit) geometries. In this case, the interface-sensitivity is achieved through a decrease in effective penetration depth. At shallow angles $\alpha$ between the surface tangent and the incoming beam, the X-ray penetration depth decreases with $\sin(\alpha)$. The absolute values of the penetration depth can then be calculated based on the material-specific and energy-dependent X-ray absorption coefficients, as tabulated by Henke et al.[123] Examples include 2 nm and 4 nm information depth for investigation of Pt and perovskite oxide surfaces, at 0.27 and 1°, respectively.[59,124] The instructive work by the groups of Tesche, Jooss and Risch also discusses in-depth the distortions caused by self-absorption effects and how to minimize them.[59]

For solid-solid interfaces, XAS and the associated magnetic circular or linear dichroism have been used, for example, to study the electronic or magnetic exchange across interfaces and at various temperatures and external stimuli. Thus, interfacial charge transfer across interfaces as in dye-sensitized solar cells can be visualized[125] and phase transitions or their suppression at an interface can be correlated to specific states and trapping at interfacial defects.[126]

For the solid-gas interface, operando XAS allowed new insights into electrochemically-induced phase transitions,[127] various (electro-)catalytic reactions such as the reduction of carbon dioxide to hydrocarbons[128] and gas sensing applications like tin oxide sensors.[129] The detailed information about interfacial redox processes allowed to further our understanding about fundamental reaction mechanisms[130] and even challenged the conventional beliefs about active sites for redox reactions. For example, the groups of Chueh and Bluhm found that surface oxygen anions in transition metal oxides were a redox partner for molecular oxygen during oxygen evolution and oxygen reduction reactions.[131] In these cases, instrumentation developed for near-ambient X-ray photoelectron spectroscopy, which will be discussed in more detail below, was very profitable.

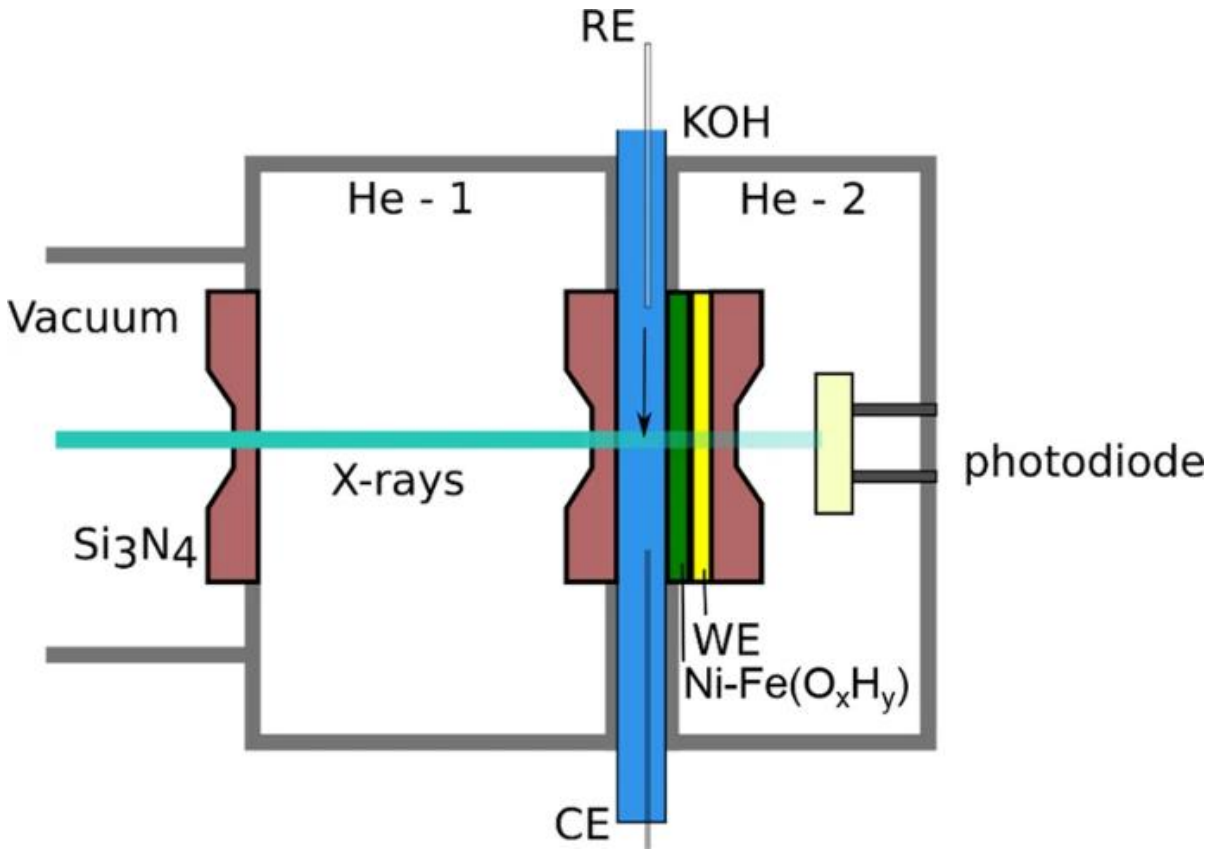

**Figure 2:** Schematic illustration of a possible geometry to measure XAS for solid-liquid interfaces. The X-rays are transmitted through several "windows", the liquid and the sample before detection with a photodiode. Reproduced with permission from [132].

For the solid-liquid interface, the experimental setup is more challenging and X-ray interaction with liquid electrolytes (formation of radicals or free electrons) is a major obstacle for correct interpretation. For solid-liquid interfaces, hard X-rays (photon energies of thousands of electron volts) are easiest to use because they do not require ultra-high vacuum (UHV) experimental chambers, allowing for comparably simple integration of a liquid cell. But many materials require investigation using also soft X-rays with typical energies of 50–1500 eV, because transition metal L-edges may be more sensitive to the oxidation state of the catalyst than their higher-energy K-edges and light elements like Li, C, N and O only have absorption edges in the soft X-ray regime. Soft X-rays have a shorter penetration depth (on the order of 1 µm) compared to hard X-rays, necessitating more complex experimental approaches ideally using thin liquid and solid layers, as shown in Figure 2 (refs. [132,133]) and as discussed in more detail by the group of Salmeron.[7] The application of operando XAS for electrocatalysts[40], solar energy materials,[96] and lithium-ion batteries[38] has been reviewed recently, so I will only give very brief examples here.

For electrocatalysts, the phase of the crystalline bulk and the oxidation state and coordination of the active sites can be mapped as a function of applied potential.[58,134] It proved extremely useful to derive reaction mechanisms based on the structure and oxidation state under reaction conditions, e.g. for organic or inorganic $CO_2$ reduction or oxygen reduction and oxygen evolution electrocatalysts.[58,119,132,134–137] For lithium-ion batteries, operando XAS has helped unravel the mechanism of the charging and discharging processes and to obtain information about intermediate phases forming during operation, as shown for example in the work by Tromp and Gasteiger.[138] Further examples can be found in refs. [77] and [95].

To summarize, the development of dedicated synchrotron endstations for operando X-ray spectroscopy for various interfaces has led to tremendous insights into the oxidation state and local geometry of active materials under operating conditions for a wide range of applications. Particularly, for electrochemical energy conversion and storage, XAS is an invaluable tool, and recent efforts even enabled laboratory-based experiments.[139] For interface sensitivity, XAS special detection modes or additional experimental protocols have to be used to pick out small spectral changes in thin interfacial layers, as will be discussed in detail in section 4.2

## 3.3 Soft and hard X-ray photoemission spectroscopy

X-ray photoelectron spectroscopy (XPS) is a versatile tool for the determination of the electronic and chemical states and the stoichiometry.[62] A detailed description of XPS can be found in several detailed review articles[64–66] and the reader is also referred to the excellent practical guidelines presented by Baer *et al.*,[67] Powell,[68] Chambers *et al.*,[69] Tougard[70] and others in a recent tutorial series of the American Vacuum Society. Here, I will focus on the essentials necessary for the discussion of interface characterization with XPS. Like XAS, XPS is based on the absorption of photons by electrons. If the photon energy is higher than the energy difference between core level and vacuum level, the excited photoelectron escapes into vacuum and can be detected. The electron kinetic energy $E'_{\mathrm{kin}}$ after leaving the sample is related to the binding energy of the initial core level $E_{\mathrm{bin}}$ according to

$$E'_{\mathrm{kin}} = h\nu - E_{\mathrm{bin}} - \Phi_{\mathrm{sample}} \quad (1)$$

with the photon energy $h\nu$ and the sample work function $\Phi_{\mathrm{sample}}$. This equation can be rewritten as

$$E_{\mathrm{kin}} = h\nu - E_{\mathrm{bin}} - \Phi_{\mathrm{analyzer}} \quad (2)$$

with the electron kinetic energy as measured by the analyzer $E_{\mathrm{kin}}$ and the analyzer work function $\Phi_{\mathrm{analyzer}}$,[64] a quantity that can be easily calibrated. So the element-specific binding energy of the core level electrons can be determined from the measured kinetic energy of the photoelectron, allowing for determination of the valence state and electronic structure. The integrated intensities of the characteristic peaks are a measure of the relative atomic concentrations after normalization with relative sensitivity factors that account for differences in the cross sections for the photoelectric effect for different elements, orbitals and instrument geometries.

X-rays can penetrate the sample and excite photoelectrons from a depth of several hundred nanometers. However, as the generated photoelectrons propagate to the surface, they scatter elastically and inelastically. Therefore, electrons which are created near the surface have a higher probability of leaving the sample with their characteristic energy and inelastically scattered electrons contribute to the background of the spectrum. Accordingly, XPS is a surface sensitive technique and the attenuated intensity $I(t)$ of photoelectrons generated at a depth $t$ can be described (in an overly simplistic picture and neglecting elastic scattering) according to:

$$I(t) = I_0 \exp \frac{-t}{\lambda_\mathrm{i} \cos \theta} \qquad 3$$

Here, $I_0$ is the photoelectron intensity without attenuation, $\theta$ is the photoemission angle (measured between the surface normal and the detector), and the inelastic mean free path $\lambda_\mathrm{i}$ is the "average distance that an electron with a given energy travels between successive inelastic collisions."[68] A more complete description was provided by Powell.[68] $\lambda_\mathrm{i}$ depends on the electron kinetic energy (and therefore on the chosen X-ray energy and the binding energy of the core level, see equation 1), as described in the NIST databases (National Institute of Standards and Technology) and predicted using the so-called TPP-2M formalism.[68] For most materials, $\lambda_\mathrm{i}$ has a minimum of ~3 to 4 Å at around 50 eV, with an increase towards higher and lower energies due to decreasing scattering rates, and the underlying mechanisms and the dependencies on material properties and experimental geometries are still subject of intense research.[68,140–142] In practice, it is useful to define a mean escape depth as a measure for the surface sensitivity of an XPS experiment with a given material and instrument configuration. Neglecting elastic scattering again, the mean escape depth is defined as:

$$\Delta = \lambda_\mathrm{i} \cos \theta \qquad 4$$

So $\Delta$ can be considered as the average depth from which the detected photoelectrons originate in the given experiment, and $\Delta$ typically has values between 0.3 nm and 2 nm for soft X-ray excitation. For hard X-ray photoelectron spectroscopy (HAXPES), $\Delta$ can be up to ~10 nm. These considerations show that the choice of photoemission angle and excitation energy is decisive for the depth sensitivity (in principle non-destructive depth-profiling is possible), and it will be shown in detail in section 4.3 that it is also decisive for the depth selectivity for interface characterization.

Because of the short inelastic mean free path of photoelectrons, XPS instrumentation typically requires UHV conditions. Recent developments now also allow characterization in other environments, as discussed in detail below. The classical UHV setups only allows investigation of solid surfaces or solid-solid interfaces close to the sample surface. Examples for the application of XPS, for example for the study of thin films[66] and numerous energy and information technologies[62] are ubiquitous and the reader is referred to the recent review articles. In situ analysis revealed, for example, charge transfer across solid-solid interfaces,[143] spin states in magnetic tunneling barriers[62] and band alignments.[66,144] Recently, the group of Dittmann used operando HAXPES to study the switching mechanism in memristive devices with thin tunneling barriers sandwiched between an active oxide layer and a metal electrode.[25,145] For energy materials, first approaches and advances have been made in the operando XPS characterization of emerging solid-state battery technologies.[146,147]

## 3.4 (Near-)ambient pressure X-ray photoemission spectroscopy

For the solid-gas interface, XPS experimentalists faced a pressure gap between the UHV operating pressures ($p < 10^{-9}$ mbar) and the relevant pressures where sensor or electrocatalysis applications operate (few mbar to pressures exceeding atmospheric pressures). Therefore, so-called ambient pressure XPS (APXPS, also referred to as near-ambient pressure XPS, NAP-XPS)) tools were developed at the beginning of the century, following the approach originally explored by Siegbahn et al. in the 1970s and 1980s,[148,149] who used differential pumping stages that progressively reduce the pressure. This is necessary to separate the high pressure near the sample from the required vacuum in the electron analyzer, to minimize the scattering probability for the electrons traveling through the high-pressure region and to prevent arcing in the electron analyzer at elevated pressure. The design was perfected using differentially-pumped analyzer lens system at the Advanced Light Source in Berkeley and at BESSY in Berlin, allowing tens of mbar operating pressures.[74,150–152] Technically, APXPS has not yet achieved operation in atmospheric pressures With further development and even commercial availability of laboratory-based setups,[153,154] APXPS has become a major trend in surface science.[44]

Extensive summaries of APXPS for the investigation of the chemical and electronic structure at solid-gas interfaces is provided in [44,71–74]. Recent examples include the mechanistic understanding of catalytic CO oxidation and Fischer-Tropsch synthesis[35,155–158] and electrochemical oxygen reduction and oxygen evolution[130,131,159,160] on various surfaces, identification of adsorbed species in PEM fuel cells,[161] the determination of the work function of nanomaterials,[162] and the $p_{O2}$ dependent defect chemistry including precipitation and formation of vacancies.[30,163] Interestingly, recent reports by Gunkel and coworkers used APXPS to show that even the defect chemical and electronic properties of buried solid-solid interfaces such as $LaAlO_3/SrTiO_3$ can be tuned and understood based on reactions at the nearby sold-gas interface.[164]

## 3.5 Meniscus XPS

The development of APXPS also enabled the XPS-based characterization of the solid-liquid interface, a new research direction that became very popular in the past few years due to the pioneering work at the Advanced Light Source in Berkeley and rapid installation of dedicated instrument endstations at several synchrotron facilities around the world which are currently being constructed or have just been commissioned.[79,165]

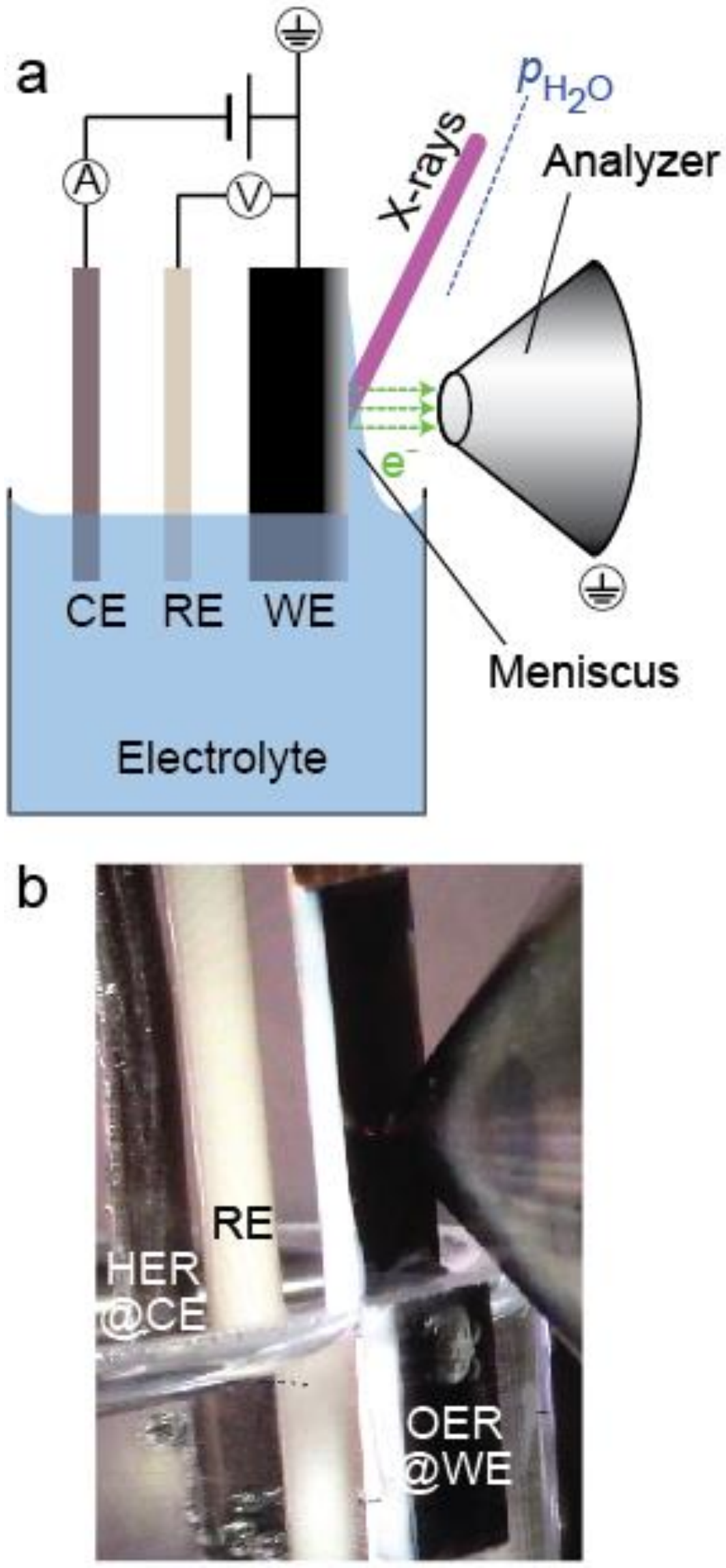

**Figure 3:** a) Schematic representation of the Meniscus XPS concept. The working electrode (WE), reference electrode (RE) and counter electrode (CE) are immersed in the aqueous electrolyte located in a APXPS chamber with a background water vapor pressure. Electrons generated at the solid-liquid interface penetrate the liquid meniscus and are collected by the analyzer cone. b) Photograph of an exemplary experiment. Anodic polarization of the WE drives the oxygen evolution reaction (OER) while cathodic polarization at the CE drives the hydrogen evolution reaction (HER). Gas evolution can be seen by the bubbles at both electrodes. Picture taken at beamline 9.3.1 of the Advanced Light Source.

If a hydrophilic solid is partially immersed in an aqueous electrolyte, a stable meniscus may form, as already discussed extensively by Bockris and Cahan.[166] In the "dip-and-pull" approach developed by Zhi Liu's groups also referred to as (hanging) meniscus XPS, the electrolyte thickness on a Pt electrode was in the range of 10 nm to 30 nm, as described in the seminal paper by Axnanda, Crumlin et al.[75] Such a thin liquid layer can be penetrated by photoelectrons. This opens up investigation of the solid-liquid junction, to study electrochemical processes like specific adsorption of ions, charge transfer dynamics and electrical potential formation.

In the "dip-and-pull" approach, a meniscus of the liquid electrolyte is obtained by immersing and partially extracting the sample from the liquid solution in a controlled ambient, as shown in Figure 3 and as described in detail in refs. [75–79]. If the partial pressure of the solvent in the chamber (in many cases: the water partial pressure) equals its vapor pressure for the experimental temperature, a stable meniscus thickness can be achieved, explaining why meniscus XPS necessitates an APXPS chamber. Alternative and also promising geometries like the "tilted sample"[167] and the "offset droplet" method using a fine capillary[168] might offer advantages regarding the proximity of the "bulk liquid" but will not be discussed here.

XPS investigation of the solid-liquid interface generally allows probing the chemical composition, the oxidation state and built-in electrical potentials via the detection of rigid photoelectron kinetic energy shifts. Accordingly, it has been applied to reveal the nature of the electrochemical double layer at electrode-electrolyte interfaces,[5] and the band alignment in photoelectrochemical cells.[169,170] Further, it was used to experimentally probe the theoretical predictions of the thermodynamically stable phases of an electrode as a function of the applied potential (so-called Pourbaix diagrams). It was found, for example, that Pt oxidation occurred at hundreds of mV higher potentials in the experiment compared to the prediction from the Pourbaix diagram.[171] Such insights are necessary to reveal the true active (surface) phases of various electrochemical materials under operating conditions to finally understand the charge transfer mechanisms, and meniscus XPS studies already contributed to such an understanding for Pt electrodes during water electrolysis used as both cathode[172] and anode[171] and for transition metal oxide based electrocatalysts for the oxygen evolution reaction.[173–175]

Despite these great advantages, there are also considerable limitations for the investigation of the solid-liquid interface through a thin meniscus. Due to mass transport limitations, the thin meniscus layer leads to at least three times higher electrolyte resistance than in the bulk, even for high electrolyte concentrations.[171] This means that investigation of the solid-liquid interface should be limited to low current densities (~below 1.0 mA $cm^{-2}$)[171] that only lead to negligible IR drop, because higher currents lead to large and ill-defined potential drops across the electrolyte. In other words, the interface under investigation might be experiencing a different potential than expected. Another related uncertainty stems from possibly non-uniform potential along the length of the electrode, as the potential drop at the solid-liquid interface likely depends on the thickness of the electrolyte layer.[7] Therefore, the applied potential should always be checked based on the relative position of electrode and electrolyte core level binding energies. Their energy difference should shift proportionally to the applied potential.[5]

In addition, the liquid layer can show instabilities because of several reasons. For example, it may change under the influence of gravity, due to slow but constant loss of electrolyte in a backfilled chamber or because of higher relative pumping in close proximity of the energy analyzer cone (which on the other hand is necessary to minimize the path length that photoelectrons have to travel through the near-ambient pressure chamber). Unfavorable potentials lead to shrinking of the stable meniscus,[166] and many faradaic reactions of interest involve consumption of the electrolyte.[176] Therefore, Stoerzinger et al. suggested and demonstrated stabilization of the meniscus through the addition of non-interacting salt.[78]

Lastly, X-ray damage or radicals created during water radiolysis must be considered in all X-ray based techniques.[77,177] These effects strongly depend on the cell design and beam energy, intensity and size.[178] Beam effects must be taken especially seriously for modern high-flux beamlines. This aspect will be addressed in the next section as well, as cell designs allowing for replenishing of the electrolyte promise to alleviate the effects to some extent.

## 3.6 Thin membranes and spectromicroscopy

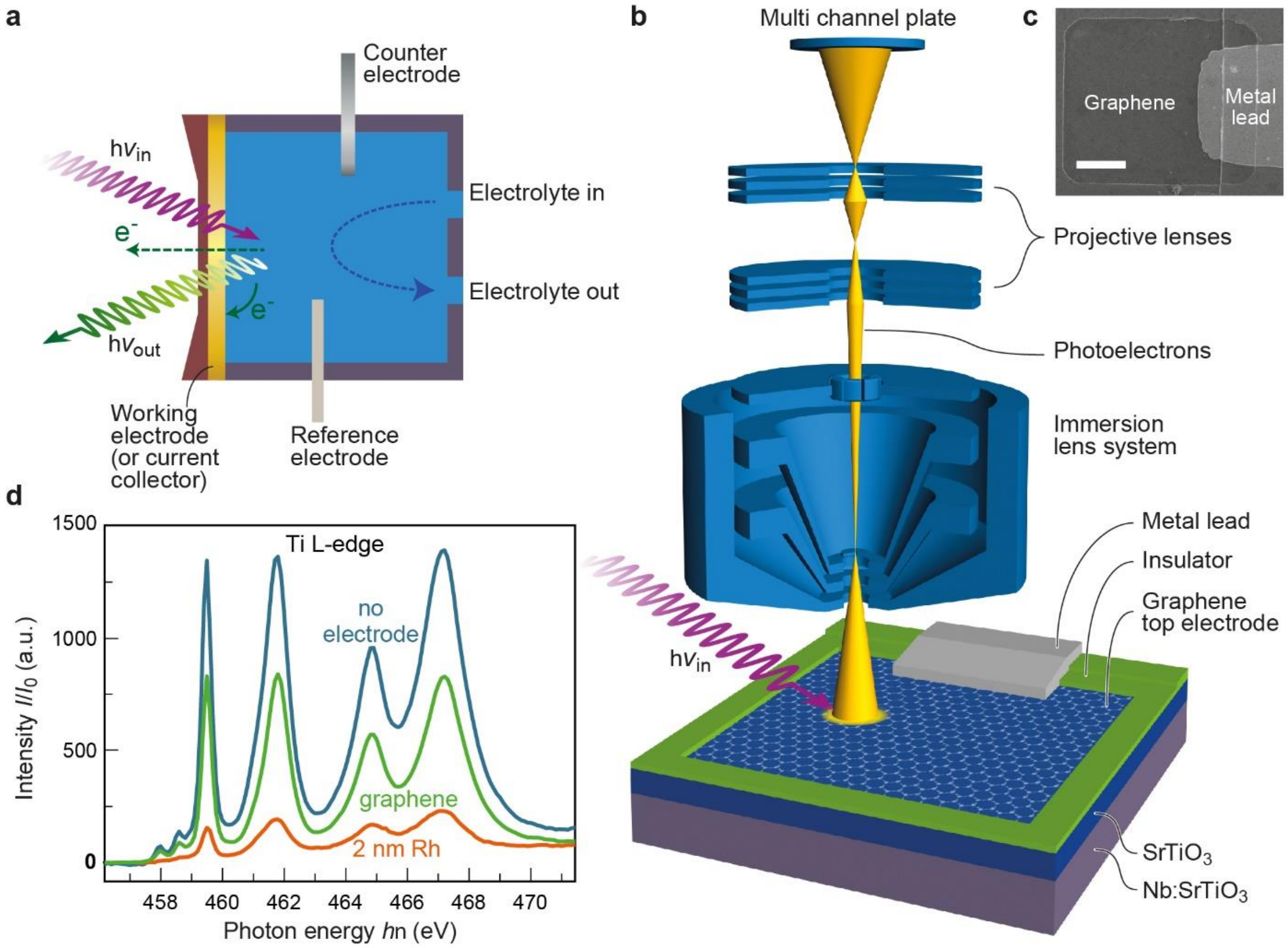

**Figure 4:** a) XAS with a thin membrane covered with a thin Au electrode.[6] XAS can be measured from fluorescence yield ($h\nu_{out}$), or by collecting secondary electrons at the thin electrode. b) Schematic of the measurement set-up for spectromicroscopy through a graphene membrane for a solid-solid interface. Here, the example is a memristive device with a $SrTiO_3$ active layer (blue) and graphene top electrode (grey honeycomb lattice). The graphene electrode is contacted through a metal lead, which is electrically separated from the continuous bottom electrode, allowing for operando biasing. At the same time, photoelectrons from the buried layer can easily escape through the graphene electrode, allowing simultaneous imaging. c) Scanning electron microscopy image of an exemplary device. Scale bar, 5 µm. d) $SrTiO_3$ Ti L-absorption edge measured without electrode, with a graphene electrode, and with a 2 nm Rh electrode. Panels b-d reproduced from [81].

An alternative approach to study solid-gas and solid-liquid interfaces is the use of thin membranes, pioneered by the groups of Salmeron[6,7,43,179,180] and Kolmakov.[181,182] This approach either uses membranes that are on the same order as the electron inelastic mean free path for XPS, or somewhat thicker membranes for XAS (typically $Si_3N_4$).[183] In contrast to meniscus XPS, the sample is illuminated and probed from the solid side, rather than the liquid side. Fluorescence photons penetrate the membrane and can be detected, leading to an XAS signal with large information depth (Figure 4a). Alternatively, XAS electron yield can be measured using a conducting layer on the solid side of the interface. In this detection mode, electrons created during the X-ray absorption and de-excitation processes in the liquid in the vicinity of the solid-liquid interface cross the interface and are detected as a current at the solid electrode, such as the Au layer used in the example shown in Figure 4a.[6] XPS can be measured in a similar manner, if the membrane is thin enough (usually graphene or graphene oxide is used). In this case, the membrane is transparent for photoelectrons, allowing detection of XPS signal from the solid-liquid interface. The details of the membrane approach to operando spectroscopy have been discussed in recent reviews and perspectives[43,48,77] and will not be discussed in great detail here.

Interface-sensitivity can be obtained if electrons are used as detected species, resulting from their limited inelastic mean free path. Successful example applications include the study of the bonding structure and orientation of water molecules at electrode interface[6] and the investigation of fundamental processes involving adsorbed species like dissociation and migration.[184] Recently, graphene membranes were combined with polymer electrolyte membranes to enhance stability of confined solid-liquid interfaces.[185] Because the electrons do not have to penetrate the liquid, the membrane-approach offers the possibility to use a flow cell setup to avoid limitations from mass transport and to decrease the effects of radicals created in water radiolysis. So the interface can be probed under more dynamic conditions. Additionally, the pressure range for solid/gas interfaces that can be probed is larger than for conventional APXPS, and it is especially attractive to combine both approaches using membrane-sealed cells in a APXPS setup.[80] Drawbacks of the technique include the limitation to selected materials and geometries due to the necessity of very thin membranes, which are difficult to prepare and prone to fail under illumination, exposure to pressure differentials or electrochemical reactions during biasing and resulting bubble formation.[43,48,80]

Another advantage of the membrane approach is that it enables interface-sensitive spectroscopic investigation with spatial resolution. For example, we used graphene as an electron-transparent membrane to study the solid/solid interface in memristive devices introduced in section 2.1. In this setup, the graphene membrane served as a top electrode for the memristive device, allowing in situ observation of spatially confined changes in the chemical and electronic properties of the buried $SrTiO_3$ active layer (Figure 4b-d).[81,186] Spectromicroscopy with graphene electrodes is very attractive because phenomena like resistive switching occur in reduced dimensions. In this case, the chemical changes are confined to small filaments in the lateral direction (typically tens to a few hundred nanometers), and to the interface between the electrode and the active layer (1-5 nm), requiring interface-sensitivity and spatial resolution simultaneously, which we achieved using photoemission electron microscopy. Figure 4d shows that the signal from the buried layers is only moderately attenuated by the graphene layer, while even 2 nm metal top electrodes lead to strong attenuation, confirming the suitability of graphene membranes for electron transparency and the interface-sensitivity of the approach.

A similar approach to study the solid-liquid interface with spatial resolution was developed and reviewed by Kolmakov and Nemsak.[82–84] This approach adds spatial resolution to the membrane-based XPS and XAS investigation of the solid-liquid interface. It is particularly attractive to thus correlate lateral inhomogeneities in electrochemical activity to variations in either chemical compositions or in the applied electric field, which can be probed directly from the relative shift of the photoelectron spectra.

### 3.7 Short summary of complementary techniques

In addition to the spectroscopic techniques discussed in this perspective, there is a multitude of complementary techniques that can provide valuable information about charge transfer processes at interfaces. For example, scanning probe techniques[89–94,187] can reveal the local double layer potential[188], add spatial resolution to vibrational spectroscopy[180] and track small changes in adsorption[189], and offer up to atomic resolution.[32,33,89]

The structure of the crystalline bulk of a material can be determined by X-ray diffraction and scattering. Using grazing incidence and analyzing crystal truncation rods allows interface-sensitivity. This so-called surface X-ray diffraction applied to a solid-liquid interfaces can identify the surface structure of the solid, reaction intermediates, and the structure of the electrolyte layer,[8,98–103] but complex data analysis with extensive curve fitting and prior knowledge of possible interface structures are necessary.[7] The nominally bulk-sensitive technique resonant inelastic X-ray scattering,[49,96,104] which provides detailed insights into the electronic structure can also become more interface-sensitive information using a grazing incidence X-ray beam.[59]

Highest spatial resolution imaging and spectroscopy can be obtained using the rapidly developing field of operando transmission electron microscopy (reviewed in [49]) where examples include the study of solid-solid interfaces in memristive devices[24] and solid-state batteries[105] and for electrochemical solid-liquid interfaces.[106–109]

## 4 Prospects for further advances in operando analysis of solid-liquid interfaces

### 4.1 Developing multiprobe experiments

The operando analysis of solid-liquid interfaces has matured in recent years and promises to deliver much needed insights into true active surface phases for various well-established and emerging technologies. But each approach still has unique experimental challenges and achieving chemical and interface-sensitivity (especially for buried interfaces) in a single method remains difficult. Based on the strengths and weaknesses of all the techniques discussed above and in the extensive literature, it is evident that a significant step forward could be achieved through the combination of multiple techniques.

When considering that multiple charge transfer processes involving the surface or the bulk often coexist or compete with each other, it is attractive to combine bulk-sensitive probes with more interface-sensitive probes. To understand and optimize future electrocatalysts, researchers require information about both the bulk and the surface, as the bulk composition crucially affects the electronic conduction and therefore the total resistance of the entire process, while the electrode surface chemical and electronic structure determine the electrocatalytic mechanisms. Obtaining this information allows synergistic tuning of bulk and surface properties to optimize the overall performance. For example, bulk-sensitive (fluorescence yield) XAS could be measured simultaneously with electron-yield XAS[6] or with meniscus XPS to probe changes in the bulk oxidation state and possible ion (de)intercalation processes and electrocatalytic reactions at the electrode surface.[175] Similar combinations of bulk and interface-sensitive probes could also be advantageous for structural probes like (surface) X-ray diffraction: X-ray diffraction can probe the bulk structure and defect chemistry,[190] while surface X-ray diffraction is a uniquely powerful tool for the interfacial structure.[102]

In addition to investigating the surface and the bulk simultaneously, the combination of dissimilar or complementary techniques is an exciting avenue to better understand various processes at the solid-liquid interface, as has already been done by combining X-ray spectroscopies and scattering techniques to probe the structure and chemistry of oxygen evolution electrocatalysts,[137,191] advocating for additional development and implementation of dedicated synchrotron endstations for such combinations. A promising example is the new apparatus combining APXPS with grazing incidence X-ray scattering at ALS, to simultaneously probe surface structure and chemistry of various interfaces under ambient conditions.[192] Combining XPS with XAS is attractive to probe the occupied and unoccupied parts of the electronic structure.[139] It is also attractive to combine X-ray based techniques with vibrational spectroscopies like infrared, UV-Vis and Raman spectroscopy, because they yield complementary information. In this sense, vibrational spectroscopy is useful to study intermediates and interfacial (adsorbed) species on the liquid side of the interface, while X-ray spectroscopy and scattering techniques reveal details about the solid side. Therefore, a given interface should be investigated using both types of techniques for a full understanding. In the ideal case, this should even be performed simultaneously. But the integration of both types of probes and detectors into a single experiment will remain challenging, so consecutive or parallel experiments of the same materials might be the best compromise.[193,194] Zhu et al. nicely reviewed the extent of understanding that can be achieved by applying multiple complementary techniques to a single system of interest.[45]

A particularly promising set of complementary techniques involves probing the solid/liquid interface with spatially averaging spectroscopic techniques of a macroscopic interface under reaction conditions with operando transmission electron microscopy, which is limited to model approaches with thin, electron-transparent specimen but which allows for atomic-resolution imaging of the surface structure, even including dynamic motion of adatoms at the solid/liquid interface.[106] Such a combination would allow derivation of atomic-level interface/structural evolution. Carbonio et al. suggested to combine X-ray spectroscopy with on-line detection of reaction products, e.g. through gas chromatography, liquid chromatography, differential electrochemical or mass spectrometry and first reports on the successful combination have been published.[77,195] Similarly, the recent successes of on-line inductively coupled plasma mass spectrometry[115,196,197] may stimulate the integration of such probes with spectroscopic experiments. Again, it is a reasonable first step to pursue such integration through clean transfer of model interfaces from one experiment to the next,[115] but in the long run, simultaneous characterization is preferable because of the dynamic changes expected.

## 4.2 Optimizing the interface-sensitivity

For the investigation of all interfaces discussed here, a critical question is how sensitive the acquired data is for the interface under investigation. While some techniques are generally bulk-sensitive but can be turned more interface-sensitive using dedicated detection modes like in XAS, other intrinsically interface-sensitive techniques like XPS (and interface-sensitive detection modes in XAS) suffer from poor signals for interface-sensitive measurements, resulting in a trade-off between interface-sensitivity and signal intensity.

Let us now discuss the example of XAS in more detail. We imagine a metallically conductive oxygen evolution electrocatalyst electrode such as $LaNiO_3$, with a flat surface exposed to a liquid electrolyte investigated with fluorescence yield to make use of the high penetration depth (Figure 5). For this Gedankenexperiment, the XAS signal can be detected in reflection or transmission geometry. Typical electrocatalyst thicknesses are tens to hundreds of nanometers, and recent reports from our group and others showed that changes in the top one to two unit cells (~0.4 to 0.8 nm) are expected during operation of $LaNiO_3$ electrocatalysts.[37,198] So the part of the electrode we are most interested in is a thin surface layer on a thick layer of the same nominal composition. For a representative 20 nm thick electrode, the signal contribution of this surface layer would therefore be ~2-5 %, assuming negligible X-ray attenuation in such a thin layer. Further, the chemically-relevant changes may translate to small changes in spectral shape, such as the shift of the adsorption edge by less than 1 eV.[199] In our 0.8 nm surface-layer, such small changes are virtually impossible to detect and track as a function of applied potential due to the small contribution to the total thickness.

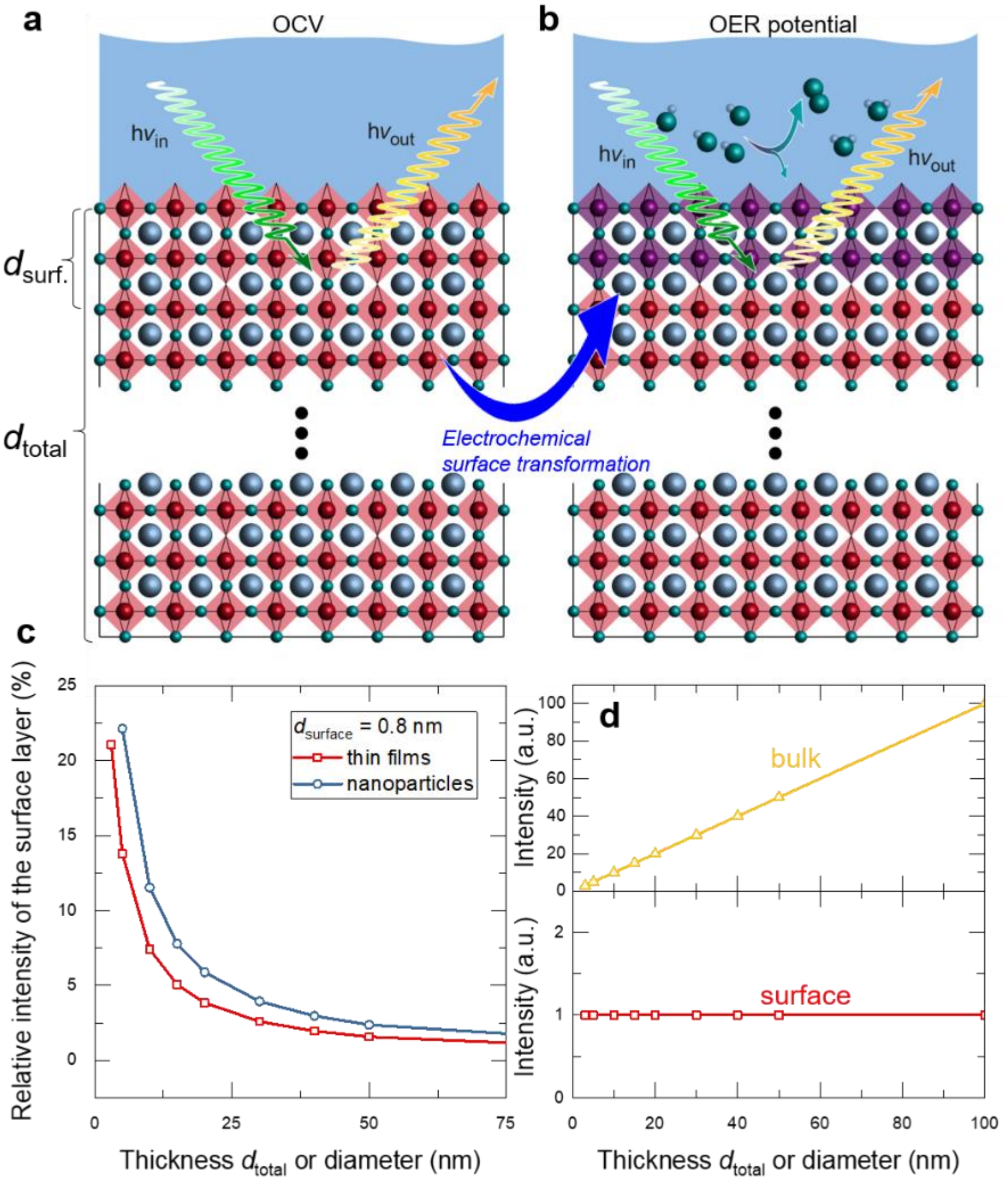

**Figure 5:** Schematic for extracting surface-sensitive information from bulk-sensitive techniques. a) Representative perovskite oxide electrocatalyst exposed to an electrolyte at open circuit voltage (OCV). b) When a potential is applied to drive the oxygen evolution reaction (OER), the same oxide undergoes a surface transformation involving the top 2 unit cells.[37] c) The relative intensity of the 0.8 nm surface layer measured by a bulk-sensitive technique like XAS or UV-Vis spectroscopy increases with decreasing total thickness of the electrode. Derived from the volume ratio of a surface-layer shell on a spherical nanoparticle. d) Bulk and surface layer intensity as a function of thickness. Using epitaxial thin films with unit-cell thickness control, one can extract the surface signal from a series of operando experiments with various thicknesses. The same dependence can be derived for nanoparticles of various diameters.

An option to overcome this limitation is to increase the relative surface area compared to the volume of the electrode material. It was suggested before that high-surface area materials like nanoparticles are an attractive pathway,[77] because they increase the relative signal intensity of the surface. Alternative pathways are the creation of highly porous 3D materials that maximize the ratio of atoms at the interface and in the bulk. For nanoparticles, and continuing the example of a thin surface layer of interest, one can estimate the contribution of the surface from the relative volume of a given sphere and the volume of a 0.8 nm spherical shell (Figure 5c). The spectral contribution is bigger than 10 % for nanoparticles with a diameter of 10 nm or smaller. A signal contribution of 10 % can be taken as a possible detection limit to still investigate the surface processes, but the exact value will depend on the exact experimental setup and the expected spectral changes. Operando spectroscopy during water electrolysis has already been performed for such small nanoparticles, so this pathway is a worthwhile avenue.[195] But the question remains how the small surface contribution will be extracted from the total signal to interpret the results quantitatively.

Recently, we therefore developed an approach to tune nominally bulk-sensitive techniques interface-sensitive using epitaxial thin films of various thickness based on the idea of Liang, Chueh and coworkers.[37] Epitaxial thin films are layers with exceptionally low defect concentration, deposited on single crystalline templates, allowing for a direct control of properties like crystalline orientation and atomic-level surface structure. The investigation of epitaxial perovskite thin films for electrocatalysis applications gained attention as a platform of tunable, well-defined electrocatalyst surfaces.[200–202] Epitaxial thin films are typically deposited with unit cell thickness precision, allowing to tune the relative contribution of the surface layer in a bulk-sensitive approach through the selection of the total thickness, as shown schematically in Figure 5a-b. Figure 5c shows the relative contribution of a two-unit-cell surface layer to the total signal for an exemplary perovskite oxide to continue our example of a 0.8 nm surface layer. Similar to the nanoparticle-approach, the spectral contribution of the surface layer is bigger than 10 % for films with a thickness below 7 nm. The relative spectral contribution of the surface layer can be increased from ~1.5 % to 14 % by decreasing the film thickness from 50 to 5 nm. Here the question arises what advantage thin films may offer compared to the nanoparticle approach discussed above. The answer lies in the high precision of epitaxial growth for model systems. By depositing the layers unit cell by unit cell, multiple samples with nominally identical properties but different thicknesses can be produced, in turn allowing for a deconvolution of spectral contributions arising from the surface and from the bulk: The surface contribution is independent of film thickness, so it should be identical for each experiment, while the bulk signal is proportional to the thickness (Figure 5d). One can therefore explicitly extract the surface-sensitive information from a series of bulk-sensitive experiment. We recently demonstrated this pathway using bulk-sensitive UV-Vis spectroscopy in transmission geometry, revealing a phase transition of a unit-cell-thin layer to a Ni hydroxide-like layer occurs during operation of a $LaNiO_3$ electrocatalyst.[37] A similar approach was also used recently to separate the processes occurring at a solid/solid interface and a solid/gas interface in close proximity.[164]

This approach evidently allows extraction of subtle changes in a very thin surface layer and can in principle be applied to all characterization tools discussed here. But of course, it also exhibits limitations: beyond the time-consuming need to repeat the same experiment with multiple thicknesses, the thickness variation in epitaxial films also raises additional concerns: Typically, the defect structure of epitaxial films depends on the thickness, with thicker films showing more defects, especially at the surface. So great care must be taken in preparation and pre-characterization of the films. Even more concerning is the possibility of thickness dependent electronic and electrochemical properties based on the band alignment at the substrate/film interface, limiting this approach to highly conducting materials,[203] where the screening lengths are short compared to the film thickness. Lastly, the overall intensity of the signal decreases with the film thickness, so that long integration times and high-intensity probes are necessary to overcome signal-to-noise-limitations in films with satisfactory interface-to-bulk signal ratios. Nevertheless, we believe this approach will yield the long-desired interface-sensitive information for a large variety of systems.

## 4.3 Future developments and interface-sensitivity of meniscus XPS

For the recently developed meniscus XPS and related XPS techniques for the study of the solid-liquid interface, technological advancements around the world are rather dynamic, because of both the promises and challenges of the technique. The main issues that have been tackled in various approaches are the stability of the liquid electrolyte and the information depth. One can anticipate that the different approaches will be brought together synergistically. Investigation of the solid-liquid interface might become possible in laboratory-based APXPS systems, where the experimental turnaround for a specific experiment can be faster, and the issues of beam damage might be less severe (at the high cost of severely prolonged integration times).

The stability of the meniscus depends on multiple properties of the materials under investigation and on the experimental geometry. As discussed in section 3.5, even for a hydrophilic and therefore suitably wettable electrode surface, the meniscus might be unstable due to the influence of gravity and solvent evaporation. Current meniscus XPS setups attempt to alleviate these limitations by dosing solvent vapor into the analysis chamber or by introducing a second (larger) solvent container into the analysis chamber to reach a constant solvent background pressure. The newly developed offset droplet method uses an alternative approach: A high pressure liquid chromatography pump applies a constant but very slow stream of electrolyte through a fine capillary close to the region of interest to balance the evaporation rate.[168] At the same time, the sample is cooled to increase droplet stability and decrease the background pressure in the chamber to increase overall intensities.[168] These rationales might be also transferred to the meniscus XPS (maintaining the advantage of having the bulk-liquid in direct contact with a part of the sample) to further improve the meniscus stability, perhaps in parallel with the chemical strategies to stabilize the liquid layer.[78] One may envision cell designs that allow cooling of the sample and electrolyte and provide a constant replenishing of the evaporated solvent. It may also be worthwhile to revisit the tilted sample geometry[167] to facilitate a macroscopic flow cell and to solve the issue of the influence of gravity.

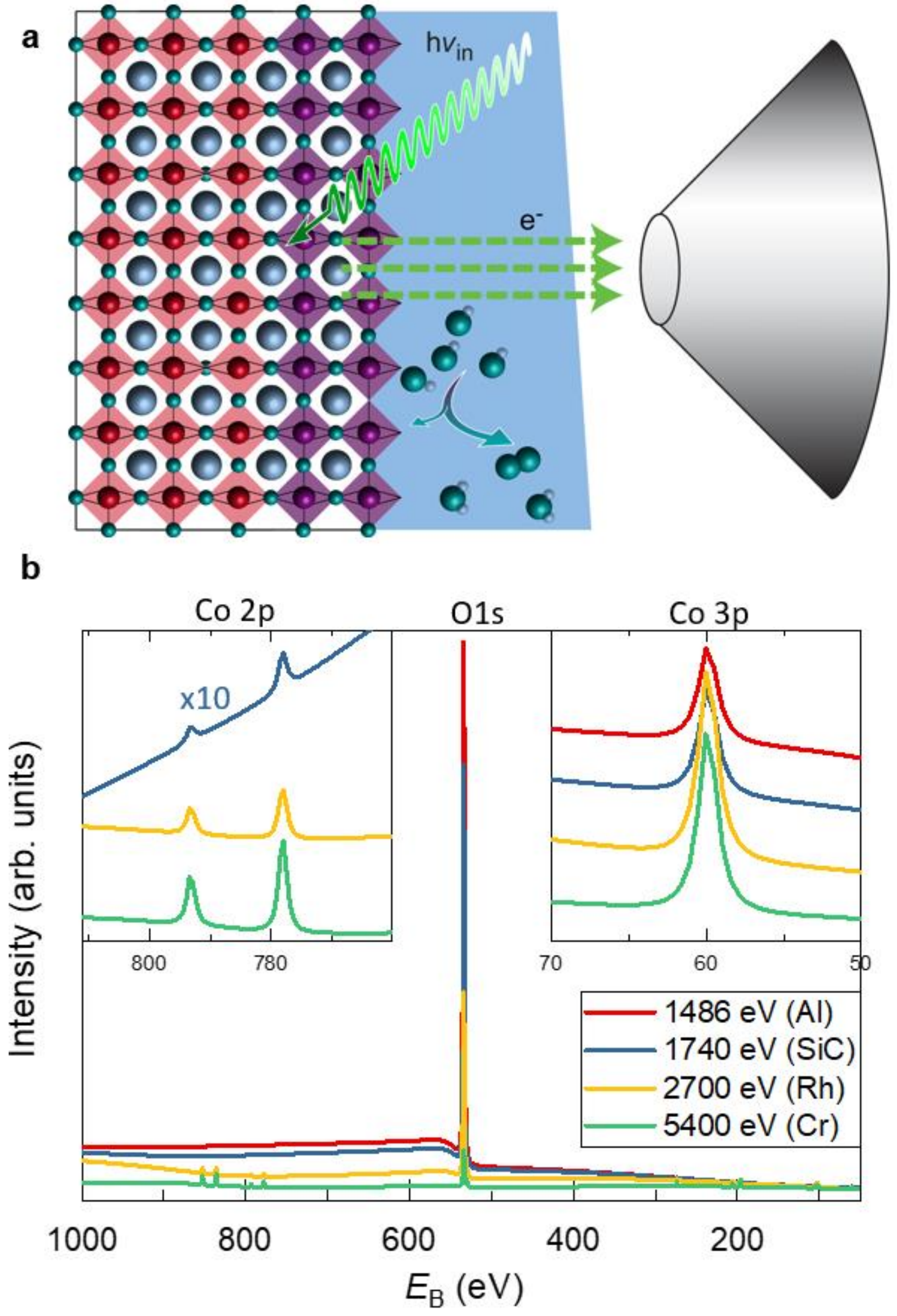

**Figure 6:** a) Schematic for SESSA simulations (not drawn to scale). A 0.8 nm $LaCoO_3$* surface layer (violet) on a 19.2 nm $LaCoO_3$ OER electrocatalyst (red) is interfaced with a 20 nm meniscus of liquid water. b) Simulated spectra. The spectra are dominated by the O 1s peak, which stems mostly from the liquid layer. The insets show a zoom-in to the Co 2p and Co 3p core levels (left and right inset, respectively). Spectra in the left inset are offset vertically for easier comparison (1486 eV is off the scale), and spectra in the right inset are displayed to-scale.

An important point in meniscus XPS (and other XPS based techniques) is the suitable selection of the X-ray energies. The very first report on meniscus XPS already included simulations performed using the SESSA software package and database (Simulation of Electron Spectra for Surface Analysis) developed by NIST.[204] It was found that "tender" X-rays with energies of around 4000 eV optimize the signal intensity of a thin overlayer on a chemically different substrate (in this case 1 nm Fe on a Si substrate) through a meniscus.[75] Later analysis showed that the ideal energy for the detection of species of the liquid side of the solid-liquid interface is also in the tender X-ray regime[60,75] and that the ideal energy also depends on the selected core level binding energy.[76] A similar analysis was also performed for illumination and detection through a thin membrane.[84]

Coming back to the electrocatalyst surface discussed in section 4.2, however, the situation is very different compared to the previous discussions in refs. [75,76]: not only the total signal of the interfacial layer but also the interface-sensitivity need to be considered. This is especially important for a thin surface layer containing the same elements as the underlying bulk of the solid. The analysis below will show that in this case hard X-rays (which are of course favorable for photoelectron penetration of the meniscus) lead to an overshadowing of the interface information by the electrons emitted from the subsurface of the solid. As electrochemical and other reactions occur at this interface and depend critically on the surface structure and chemistry, the use of such X-rays therefore impedes obtaining the relevant information, making the correct choice of the X-ray energies even more important for such systems.

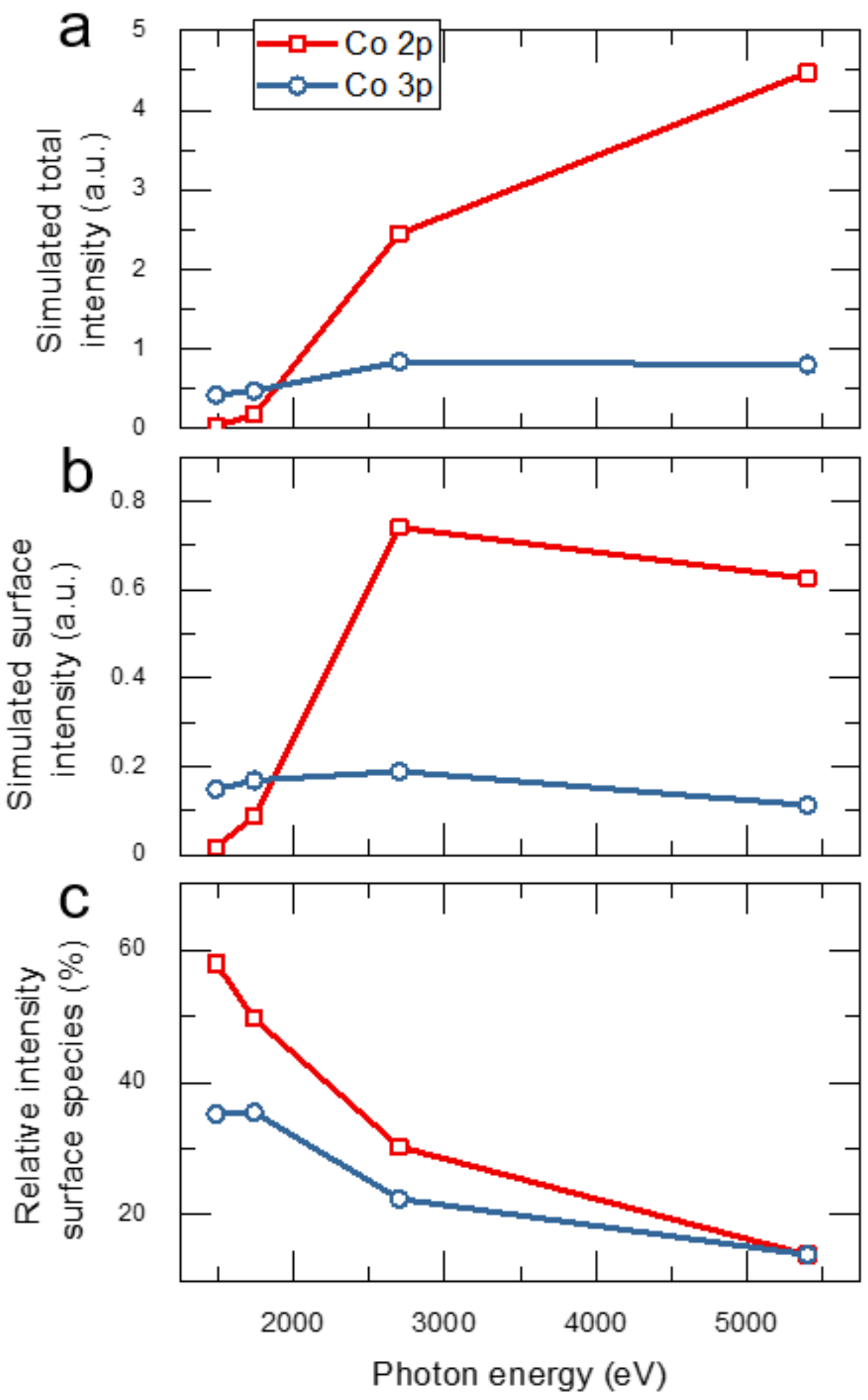

**Figure 7:** a) Total simulated intensity of the Co2p and Co 3p core levels. b) Simulated intensity of the same core levels, considering only photoelectrons from the two-unit-cell surface layer. c) Relative contribution of the surface signal to the total intensity.

To assess this situation quantitatively, SESSA simulations were performed on a system resembling the experimental setup in meniscus XPS with a $LaCoO_3$ electrode, chosen as a typical representative of a perovskite electrocatalyst without easily dissolvable species.[134,205,206] Inspired by our findings for $LaNiO_3$,[37] it is assumed that the Co chemistry of the top one to two unit cells changes as a function of applied potential, i.e. we want to identify the chemical state of Co in the top 0.8 nm.

Figure 6a shows the simulated geometry, where the 20 nm electrode is divided into a 0.8 nm $LaCoO_3$* surface layer and a $LaCoO_3$ bulk layer. Representative of an aqueous electrolyte meniscus, a 20 nm layer of liquid water layer is placed atop the sample surface. The water density and band gap used in the SESSA simulations were 1.0 g $cm^{-3}$ and 6.9 eV, as in previous simulations.[76] The inelastic and elastic scattering were simulated using the DIIMFP01 and ECS01 databases included in SESSA and the photoionization cross sections are based on the PCS01 database. The geometry was chosen to represent a typical laboratory-based experiment with the analyzer and source set at the magic angle and the sample in normal emission, with non-polarized light and an acceptance angle of 44°. Qualitatively similar trends were obtained with a typical synchrotron beamline geometry (90° angle between analyzer and detector, normal emission, grazing incidence, horizontally polarized light). Four representative X-ray energies were chosen, which can be obtained with X-ray sources available for laboratory-based experiments: Al $K_\alpha$ (1486.6 eV), SiC $K_\alpha$ (1740 eV), Rh $L_\alpha$ (2697 eV), Cr $K_\alpha$ (5417 eV). The simulated spectra are shown in Figure 6b with insets for the Co 2p and Co 3p peaks, which can be used to determine the oxidation state of Co. The simulation reveals that the total Co peak intensities increase with increasing photon energies, while the peak intensity of the O 1s and O 2s peaks, which are dominated by the signal from the electrolyte layer, and their corresponding contributions to the spectral background decrease with increasing photon energies. The increase of the Co intensities results from the increasing inelastic mean free path, while the decrease in O intensities mainly stems from a decreasing photoionization cross section.[62,76] The integrated Co peak intensities after background subtraction are shown in Figure 7a.

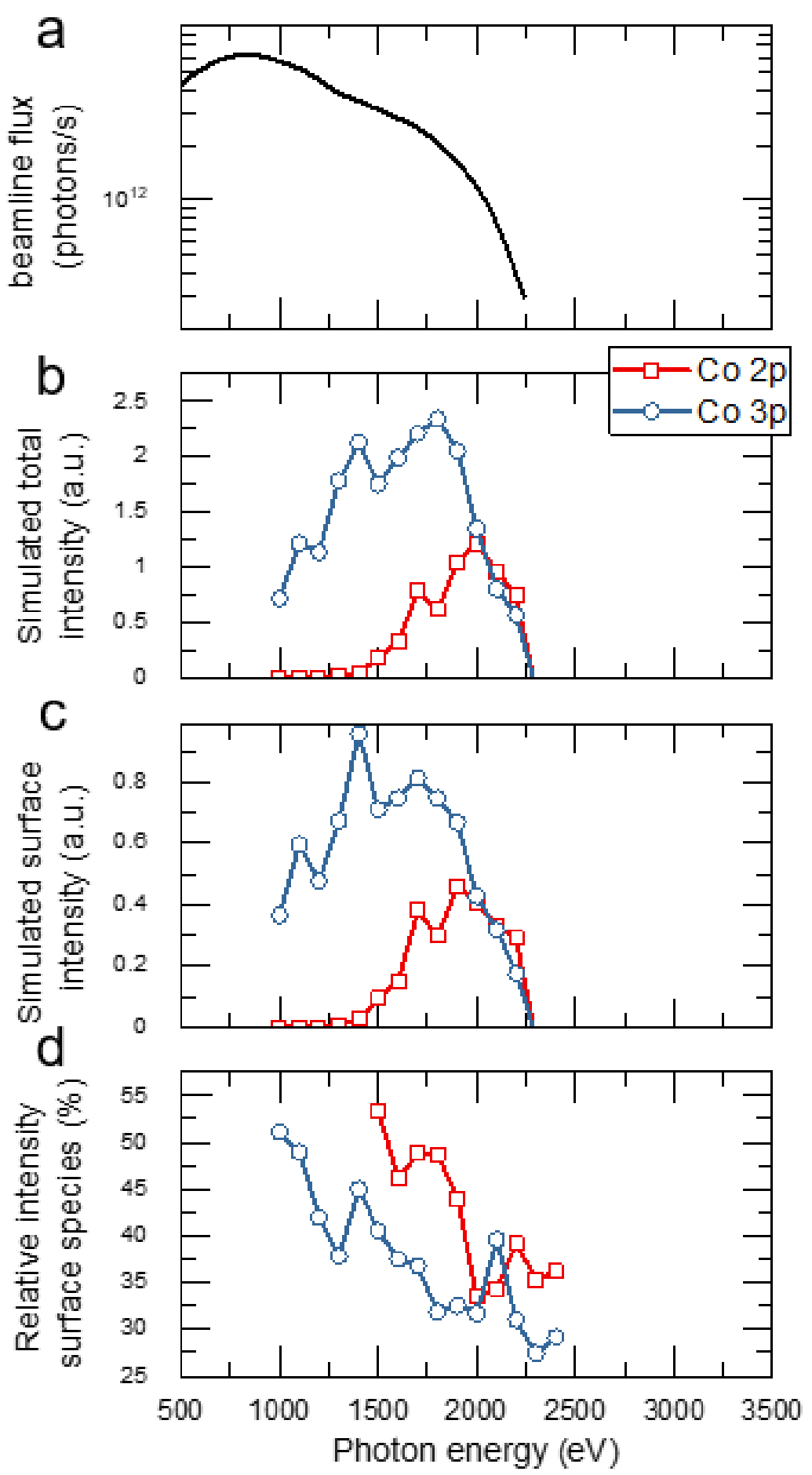

**Figure 8:** a) Beamline flux at the HIPPIE beamline at MAX IV Laboratory in Lund, redrawn from [207]. b) Total simulated intensity of the Co2p and Co 3p core levels. c) Simulated intensity of the same core levels, considering only photoelectrons from the two-unit-cell surface layer. d) Relative contribution of the surface signal to the total intensity.

To consider the interface-sensitivity, the Co peak intensities are divided into a surface species for the top 0.8 nm, where we assume the chemically relevant changes occur, and a bulk species for the underlying electrode. Figure 7b shows that the signal from the surface layer is maximized for the "tender" X-rays (2697 eV), and that the extent of intensity modulation depends on the respective core level binding energy, in agreement with the previous findings. However, the relative contribution of the surface layer, shows a fundamentally different behavior: Because of the inelastic mean free path increase, the relative surface intensity decreases monotonically with increasing photon energy (Figure 7c). The key outcome of this analysis therefore is that the *relative intensity* of the surface contribution scales differently with photon energy than the *total intensity*. To track small spectral changes originating from a thin surface layer, such as a change in Co oxidation state that is most easily recognized based on the satellite features of the Co 2p spectrum and small chemical shifts < 0.5 eV,[208] one should therefore choose the minimum photon energy that still yields sufficient total count rate. The count rate and in particular the signal-to-noise ratio depend on the available instrumentation and integration time, so no direct selection can be made for the example shown in Figure 7. For Co 2p, 1740 eV or 2697 eV may be the best compromise between total intensity and interface-sensitivity. Interestingly, the total intensity of the Co3p level is less dependent on the photon energy, so for Co 3p 1486.6 eV or 1740 eV is ideal, i.e., the typical Al $K_\alpha$ X-ray sources in modern laboratory XPS instruments are very promising.

Ideally, multiple X-ray sources (or various X-ray energies in a synchrotron experiment) should be used to carefully extract the surface information through comparison of the respective contribution to the total signal.

In addition to the cross section and inelastic mean free path considerations discussed above, the intensity of the source is of course an equally important parameter to determine the expected total intensities and signal-to-noise ratios. For laboratory sources, the soft X-ray sources are typically more intense than the hard X-ray sources, making the soft X-rays even more attractive. At the international synchrotron facilities, available APXPS beamlines with the option for meniscus XPS are optimized for distinctly different photon energies: The original instrumentation at the Advanced Light Source in Berkeley has a maximum photon flux in the "tender" X-ray range, around 4000 eV, while the recently commissioned beamline at the MAX IV Laboratory in Lund is a soft X-ray beamline. To demonstrate the importance of the beamline characteristics, the analysis from Figure 7 was repeated while considering the beamline flux at the MAX IV HIPPIE beamline. Figure 8 shows the published beamline flux and the resulting intensity profiles for the total intensity, the surface intensity, and the relative surface intensity of the exemplary $LaCoO_3$ electrocatalyst. Again, the relative contribution of the surface decreases with increasing photon energy. The absolute intensity of the Co peaks is maximized at around 1500-2000 eV, suggesting that this is the ideal energy to investigate such a system at this beamline.

Another possibility to achieve the desired interface sensitivity is turning to angle-dependent measurements,[209–212] which has achieved depth-profiling with a resolution of roughly $\frac{\lambda_i}{3}$.[210] Surface concentrations and overlayer thickness, can be extracted rather straightforwardly when varying the photoemission angle, based on equation 2. Using certain assumptions about sample properties (e.g. about the roughness and layer thickness uniformity), the main features of the concentration-depth profile can also be extracted to a fair degree of accuracy.[213] For the solid-liquid interface, however, it remains a formidable challenge to vary the photoemission angle in the required range of ~40° necessary to extract a depth profile while maintaining a stable meniscus. Another route might be the small angle variation (a fraction of a degree up to ~3°) necessary to achieve depth profiling in standing wave XPS, as discussed below.

To summarize, the XPS simulations demonstrate that for each research question, the instrument, geometry and in particular the X-ray energy must be chosen wisely, ideally based on the prior simulation of the expected intensities and interface-sensitivity. In the chosen example task of tracking small chemical changes in a thin surface layer on a chemically similar bulk electrode, the analysis revealed that it is more beneficial to use softer X-rays up to roughly 2000 eV rather than harder X-rays in the 3000-5500 eV range because of a loss of interface-sensitivity for harder X-rays. The development of laboratory tools or synchrotron endstations with X-ray energies of 1000-2000 eV are therefore highly desirable. Nonetheless, the higher inelastic mean free path of hard X-rays remains useful for many research questions which require larger information depth, or which are less reliant on interface-sensitivity. Ideally, APXPS instruments should therefore offer multiple X-ray sources of various energies to answer a diverse set of research questions. Development of instruments compatible with multiple synchrotron endstations[79] of different photon energy are therefore commendable.

## 4.4 Outlook

With increasing commercial availability and continuous efforts to develop beyond-state-of-the art instrumentation at synchrotron facilities around the world, the (spectroscopic) operando characterization of solid-solid, solid-gas and solid-liquid interfaces will continue to play a major role in cutting-edge research activities for a variety of fields, and the discussed developments will facilitate the challenging characterization of electrochemical solid-liquid interfaces for energy technologies.

The techniques discussed here can become even more interface-sensitive if they are combined with so-called standing wave XPS or XAS. In this approach, X-ray standing waves are generated at a superlattice of optically dissimilar materials. The vertical position of antinodes of the standing waves can be shifted by varying the incident angle, resulting in precise depth-selective photoemission intensity modulations. As a result, sub-unit cell depth resolution can be achieved.[62,88] For our example of a surface transformation in $LaNiO_3$ electrocatalysts, ex situ standing wave XPS already demonstrated that the chemical changes were confined to a 4 Å surface layer.[37] Few pioneering studies by the groups of Fadley, Bluhm and Nemšák applied standing wave XPS to solid-liquid interfaces and revealed great potential.[85,86] I therefore expect a larger trend in operando standing wave XPS studies for highest interface-sensitivity. An additional promising avenue to achieve this interface-sensitivity lies finely tuning the penetration depth of X-rays in near total reflection mode, as discussed nicely in ref. [214].

Lastly, it will be a general task for the community to combine interface-sensitivity with resolution in additional dimensions: 1) For spatial resolution, the use of spectromicroscopic techniques with thin membranes will be extended. The graphene membranes can also serve as electrodes in solid-state applications,[81] so emerging devices and functionalities can be explored during in situ stimulation. For example, ferroelectric switching underneath a graphene membrane, which we explored several years ago,[215] can now be probed operando. For the spectromicroscopy of solid-liquid interfaces, the development of near ambient pressure[216] or transmission[217] photoelectron microscopes can be envisioned to help overcome limitations from the challenging sample design and robustness. 2) For the time dimension, synchrotron-based pump-probe experiments[218] and ultra-fast laser setups[219,220] can yield picosecond resolution for the study of excited/intermediate states during charge generation and transfer. Alternatively, so-called “time-multiplexed” techniques may help overcome signal variations from spatial drift, changes in the background absorption, or incoming X-ray intensity to isolate weak signals during longer integration times.[221]

# 5 Summary

In this perspective, I briefly discussed X-ray spectroscopic techniques for operando characterization of solid-solid, solid-gas and solid-liquid interfaces. Recent developments now allow a wide range of possibilities, yet the interface-sensitivity remains a major concern. I described a new method to extract interface-sensitive information from nominally bulk sensitive techniques, which was developed together with Liang and Chueh at Stanford University. The method relies on precise thickness control, e.g. achievable in epitaxial growth, to repeat the same operando experiment with multiple thicknesses. For the recently developed meniscus XPS, SESSA simulations revealed that while the intensity of a solid-liquid interface might be optimized using "tender" X-rays, the interface-sensitivity is maximized for soft X-rays. This means that always a compromise between intensity and interface-sensitivity must be found, and suggests that prior to operando characterization, the expected geometry and changes under operating conditions should be simulated before designing the experiment. It will be helpful to combine multiple X-ray energies (and multiple techniques) to understand and distinguish processes at interfaces and in the bulk of a material.

In the end, the continuous development of operando probes and their combination will help us address the urgent open question regarding charge transfer reactions in numerous technologies including energy conversion and storage, sensing, manufacturing, and information technology. In many instances, we will be enabled to finally provide an answer to the question: What is the chemical composition and electronic structure across the interface of interest during operation?

# 6 Acknowledgement

The author thanks Allen Yu-Lun Liang and William Chueh for their important roles in developing the method discussed in section 4.2 and Slavomír Nemšák, Felix Gunkel, Moniek Tromp, and Gertjan Koster for fruitful discussion on interfaces and operando spectroscopy. Support of the University of Twente in the framework of the tenure track startup package is gratefully acknowledged.

# 7 AIP Publishing Data Sharing Policy

The data that support the findings of this study are available from the corresponding author upon reasonable request.

# 8 References

[1] H. Helmholtz, Ann. Der Phys. Und Chemie **243**, 337 (1879).

[2] J.F. Daniell and W.A. Miller, Philos. Trans. R. Soc. London **134**, 1 (1844).

[3] G. Jerkiewicz, in *Solid-Liquid Electrochem. Interfaces*, edited by G. Jerkiewicz, M.P. Soriaga, A. Wieckowski, and K. Uosaki (American Chemical Society, Washington, DC, 1997), pp. 1–12.

[4] O. Björneholm, M.H. Hansen, A. Hodgson, L.-M. Liu, D.T. Limmer, A. Michaelides, P. Pedevilla, J. Rossmeisl, H. Shen, G. Tocci, E. Tyrode, M.-M. Walz, J. Werner, and H. Bluhm, Chem. Rev. **116**, 7698 (2016).

[5] M. Favaro, B. Jeong, P.N. Ross, J. Yano, Z. Hussain, Z. Liu, and E.J. Crumlin, Nat. Commun. **7**, 1 (2016).

[6] J.-J. Velasco-Velez, C.H. Wu, T.A. Pascal, L.F. Wan, J. Guo, D. Prendergast, and M. Salmeron, Science **346**, 831 (2014).

[7] C.H. Wu, R.S. Weatherup, and M.B. Salmeron, Phys. Chem. Chem. Phys. **17**, 30229 (2015).

[8] M.F. Toney, J.N. Howard, J. Richer, G.L. Borges, J.G. Gordon, O.R. Melroy, D.G. Wiesler, D. Yee, and L.B. Sorensen, Nature **368**, 444 (1994).

[9] F. Gunkel, D. V. Christensen, and N. Pryds, J. Mater. Chem. C **8**, 11354 (2020).

[10] J. Mannhart and D.G. Schlom, Science **327**, 1607 (2010).

[11] H.Y. Hwang, Y. Iwasa, M. Kawasaki, B. Keimer, N. Nagaosa, and Y. Tokura, Nat. Mater. **11**, 103 (2012).

[12] J. Hoffman, I.C. Tung, B.B. Nelson-Cheeseman, M. Liu, J.W. Freeland, and A. Bhattacharya, Phys. Rev. B **88**, 144411 (2013).

[13] Z. Zhong and P. Hansmann, Phys. Rev. X **7**, 011023 (2017).

[14] A.S. Disa, D.P. Kumah, A. Malashevich, H. Chen, D.A. Arena, E.D. Specht, S. Ismail-Beigi, F.J. Walker, and C.H. Ahn, Phys. Rev. Lett. **114**, 026801 (2015).

[15] G. Araizi-Kanoutas, J. Geessinck, N. Gauquelin, S. Smit, X.H. Verbeek, S.K. Mishra, P. Bencok, C. Schlueter, T.-L. Lee, D. Krishnan, J. Fatermans, J. Verbeeck, G. Rijnders, G. Koster, and M.S. Golden, Phys. Rev. Mater. **4**, 026001 (2020).

[16] F. Gunkel, P. Brinks, S. Hoffmann-Eifert, R. Dittmann, M. Huijben, J.E. Kleibeuker, G. Koster, G. Rijnders, and R. Waser, Appl. Phys. Lett. **100**, 052103 (2012).

[17] F. Gunkel, S. Wicklein, S. Hoffmann-Eifert, P. Meuffels, P. Brinks, M. Huijben, G. Rijnders, R. Waser, and R. Dittmann, Nanoscale **7**, 1013 (2015).

[18] A. Kalabukhov, R. Gunnarsson, J. Börjesson, E. Olsson, T. Claeson, and D. Winkler, Phys. Rev. B **75**, 121404 (2007).

[19] S.A. Chambers, M.H. Engelhard, V. Shutthanandan, Z. Zhu, T.C. Droubay, L. Qiao, P.V. Sushko, T.

Feng, H.D. Lee, T. Gustafsson, E. Garfunkel, A.B. Shah, J.-M. Zuo, and Q.M. Ramasse, Surf. Sci. Rep. **65**, 317 (2010).

[20] J. Janek and W.G. Zeier, Nat. Energy **1**, 16141 (2016).

[21] Q. Zhao, S. Stalin, C.-Z. Zhao, and L.A. Archer, Nat. Rev. Mater. **5**, 229 (2020).

[22] S. Xia, X. Wu, Z. Zhang, Y. Cui, and W. Liu, Chem **5**, 753 (2019).

[23] W. Kim, S. Menzel, D.J. Wouters, Y. Guo, J. Robertson, B. Roesgen, R. Waser, and V. Rana, Nanoscale **8**, 17774 (2016).

[24] D. Cooper, C. Baeumer, N. Bernier, A. Marchewka, C. La Torre, R.E. Dunin-Borkowski, S. Menzel, R. Waser, and R. Dittmann, Adv. Mater. **29**, 1700212 (2017).

[25] C. Baeumer, T. Heisig, B. Arndt, K. Skaja, F. Borgatti, F. Offi, F. Motti, G. Panaccione, R. Waser, S. Menzel, and R. Dittmann, Faraday Discuss. (2019).

[26] C. Bäumer, R. Dittmann, C. Baeumer, and R. Dittmann, in *Met. Oxide-Based Thin Film Struct.*, edited by N. Pryds and V. Esposito (Elsevier, Amsterdam, 2018), pp. 489–522.

[27] R. Waser, R. Dittmann, G. Staikov, and K. Szot, Adv. Mater. **21**, 2632 (2009).

[28] F. Hess, A.T. Staykov, B. Yildiz, and J. Kilner, in *Handb. Mater. Model.*, edited by W. Andreoni and S. Yip (Springer International Publishing, Cham, 2019), pp. 1–31.

[29] M. Acosta, F. Baiutti, A. Tarancón, and J.L. MacManus-Driscoll, Adv. Mater. Interfaces **6**, 1900462 (2019).

[30] N. Tsvetkov, Q. Lu, L. Sun, E.J. Crumlin, and B. Yildiz, Nat. Mater. **15**, 1010 (2016).

[31] F. Fischer and H. Tropsch, Brennst. Chem **4**, 276 (1923).

[32] J. Balajka, M.A. Hines, W.J.I. DeBenedetti, M. Komora, J. Pavelec, M. Schmid, and U. Diebold, Science **361**, 786 (2018).

[33] M. Riva, M. Kubicek, X. Hao, G. Franceschi, S. Gerhold, M. Schmid, H. Hutter, J. Fleig, C. Franchini, B. Yildiz, and U. Diebold, Nat. Commun. **9**, 3710 (2018).

[34] F. Polo-Garzon, V. Fung, X. Liu, Z.D. Hood, E.E. Bickel, L. Bai, H. Tian, G.S. Foo, M. Chi, D. Jiang, and Z. Wu, ACS Catal. **8**, 10306 (2018).

[35] K. Huang, X. Chu, L. Yuan, W. Feng, X. Wu, X. Wang, and S. Feng, Chem. Commun. **50**, 9200 (2014).

[36] D.Y. Chung, P.P. Lopes, P. Farinazzo Bergamo Dias Martins, H. He, T. Kawaguchi, P. Zapol, H. You,

D. Tripkovic, D. Strmcnik, Y. Zhu, S. Seifert, S. Lee, V.R. Stamenkovic, and N.M. Markovic, Nat. Energy **5**, 222 (2020).

[37] C. Baeumer, J. Li, Q. Lu, A.Y.-L. Liang, L. Jin, H.P. Martins, T. Duchoň, M. Glöß, S.M. Gericke, M.A. Wohlgemuth, M. Giesen, E.E. Penn, R. Dittmann, F. Gunkel, R. Waser, M. Bajdich, S. Nemšák, J.T. Mefford, and W.C. Chueh, Nat. Mater. **accepted**, (2021).

[38] S.-M. Bak, Z. Shadike, R. Lin, X. Yu, and X.-Q. Yang, NPG Asia Mater. **10**, 563 (2018).

[39] M.R. Lukatskaya, O. Mashtalir, C.E. Ren, Y. Dall'Agnese, P. Rozier, P.L. Taberna, M. Naguib, P. Simon, M.W. Barsoum, and Y. Gogotsi, Science **341**, 1502 (2013).

[40] J. Timoshenko and B. Roldan Cuenya, Chem. Rev. acs. chemrev.0c00396 (2020).

[41] J. Kibsgaard and I. Chorkendorff, Nat. Energy **4**, 430 (2019).

[42] G. Zwaschka, I. Nahalka, A. Marchioro, Y. Tong, S. Roke, and R.K. Campen, ACS Catal. **10**, 6084 (2020).

[43] M. Salmeron, Top. Catal. **61**, 2044 (2018).

[44] D.E. Starr, Z. Liu, M. Hävecker, A. Knop-Gericke, and H. Bluhm, Chem. Soc. Rev. **42**, 5833 (2013).

[45] K. Zhu, X. Zhu, and W. Yang, Angew. Chemie Int. Ed. **58**, 1252 (2019).

[46] Y.-W. Choi, H. Mistry, and B. Roldan Cuenya, Curr. Opin. Electrochem. **1**, 95 (2017).

[47] M.L. Traulsen, C. Chatzichristodoulou, K. V. Hansen, L.T. Kuhn, P. Holtappels, and M.B. Mogensen, ECS Trans. **66**, 3 (2015).

[48] D.M. Itkis, J.J. Velasco-Velez, A. Knop-Gericke, A. Vyalikh, M. V. Avdeev, and L. V. Yashina, ChemElectroChem **2**, 1427 (2015).

[49] D. Liu, Z. Shadike, R. Lin, K. Qian, H. Li, K. Li, S. Wang, Q. Yu, M. Liu, S. Ganapathy, X. Qin, Q. Yang, M. Wagemaker, F. Kang, X. Yang, and B. Li, Adv. Mater. **31**, 1806620 (2019).

[50] G. Mul, F. de Groot, B. Mojet-Mol, and M. Tromp, in *Catal. An Integr. Textb. Students*, edited by U. Hanefeld and L. Lefferts (Wiley-VCH, Weinheim, Germany, 2017), pp. 271–314.

[51] P. Hollins, in *Encycl. Anal. Chem.* (John Wiley & Sons, Ltd, Chichester, UK, 2006), pp. 1–17.

[52] F.S. Gittleson, K.P.C. Yao, D.G. Kwabi, S.Y. Sayed, W.-H. Ryu, Y. Shao-Horn, and A.D. Taylor, ChemElectroChem **2**, 1446 (2015).

[53] Y.R. Shen and V. Ostroverkhov, Chem. Rev. **106**, 1140 (2006).

[54] X. Han, T. Balgar, and E. Hasselbrink, J. Chem. Phys. **130**, 134701 (2009).

[55] H. Förster, in *Charact. I. Mol. Sieves – Sci. Technol.*, edited by H.G. Karge and J. Weitkamp (Springer, Berlin, Heidelberg, 2004), pp. 337–426.

[56] B.J. Trześniewski, O. Diaz-Morales, D.A. Vermaas, A. Longo, W. Bras, M.T.M.M. Koper, and W.A. Smith, J. Am. Chem. Soc. **137**, 15112 (2015).

[57] L. Francàs, S. Corby, S. Selim, D. Lee, C.A. Mesa, R. Godin, E. Pastor, I.E.L. Stephens, K.-S. Choi, and J.R. Durrant, Nat. Commun. **10**, 5208 (2019).

[58] Y. Gorlin, B. Lassalle-Kaiser, J.D. Benck, S. Gul, S.M. Webb, V.K. Yachandra, J. Yano, and T.F. Jaramillo, J. Am. Chem. Soc. **135**, 8525 (2013).

[59] P. Busse, Z. Yin, D. Mierwaldt, J. Scholz, B. Kressdorf, L. Glaser, P.S. Miedema, A. Rothkirch, J. Viefhaus, C. Jooss, S. Techert, and M. Risch, J. Phys. Chem. C **124**, 7893 (2020).

[60] D.E. Starr, M. Favaro, F.F. Abdi, H. Bluhm, E.J. Crumlin, and R. van de Krol, J. Electron Spectros. Relat. Phenomena **221**, 106 (2017).

[61] P. V. Sushko and S.A. Chambers, Sci. Rep. **10**, 13028 (2020).

[62] M. Müller, S. Nemšák, L. Plucinski, and C.M. Schneider, J. Electron Spectros. Relat. Phenomena **208**, 24 (2016).

[63] C.S. Fadley, in *Hard X-Ray Photoelectron Spectrosc. ( HAXPES )*, edited by J.C. Woicik (Heidelberg, New York, 2016), pp. 1–34.

[64] C.S. Fadley, J. Electron Spectros. Relat. Phenomena **178**–**179**, 2 (2010).

[65] D. Briggs and J.T. Grant, *Surface Analysis by Auger and X-Ray Photoelectron Spectroscopy* (IM Publications, Chichester, West Sussex, U.K., 2003).

[66] S. Thapa, R. Paudel, M.D. Blanchet, P.T. Gemperline, and R.B. Comes, J. Mater. Res. 1 (2020).

[67] D.R. Baer, K. Artyushkova, C. Richard Brundle, J.E. Castle, M.H. Engelhard, K.J. Gaskell, J.T. Grant, R.T. Haasch, M.R. Linford, C.J. Powell, A.G. Shard, P.M.A. Sherwood, and V.S. Smentkowski, J. Vac. Sci. Technol. A **37**, 031401 (2019).

[68] C.J. Powell, J. Vac. Sci. Technol. A **38**, 023209 (2020).

[69] S.A. Chambers, L. Wang, and D.R. Baer, J. Vac. Sci. Technol. A **38**, 061201 (2020).

[70] S. Tougaard, J. Vac. Sci. Technol. A **39**, 011201 (2021).

[71] K. Roy, L. Artiglia, and J.A. van Bokhoven, ChemCatChem **10**, 666 (2018).

[72] K.A. Stoerzinger, W.T. Hong, E.J. Crumlin, H. Bluhm, and Y. Shao-Horn, Acc. Chem. Res. **48**, 2976 (2015).

[73] J. Schnadt, J. Knudsen, and N. Johansson, J. Phys. Condens. Matter **32**, 413003 (2020).

[74] A. Shavorskiy, O. Karslioglu, I. Zegkinoglou, and H. Bluhm, Synchrotron Radiat. News **27**, 14 (2014).

[75] S. Axnanda, E.J. Crumlin, B. Mao, S. Rani, R. Chang, P.G. Karlsson, M.O.M. Edwards, M. Lundqvist, R. Moberg, P. Ross, Z. Hussain, and Z. Liu, Sci. Rep. **5**, 9788 (2015).

[76] M. Favaro, F. Abdi, E. Crumlin, Z. Liu, R. van de Krol, and D. Starr, Surfaces **2**, 78 (2019).

[77] E.A. Carbonio, J.-J. Velasco-Velez, R. Schlögl, and A. Knop-Gericke, J. Electrochem. Soc. **167**, 054509 (2020).

[78] K.A. Stoerzinger, M. Favaro, P.N. Ross, Z. Hussain, Z. Liu, J. Yano, and E.J. Crumlin, Top. Catal. **61**, 2152 (2018).

[79] Z. Novotny, D. Aegerter, N. Comini, B. Tobler, L. Artiglia, U. Maier, T. Moehl, E. Fabbri, T. Huthwelker, T.J. Schmidt, M. Ammann, J.A. van Bokhoven, J. Raabe, and J. Osterwalder, Rev. Sci. Instrum. **91**, 023103 (2020).

[80] A. Kolmakov, L. Gregoratti, M. Kiskinova, and S. Günther, Top. Catal. **59**, 448 (2016).

[81] C. Baeumer, C. Schmitz, A. Marchewka, D.N. Mueller, R. Valenta, J. Hackl, N. Raab, S.P. Rogers, M.I. Khan, S. Nemsak, M. Shim, S. Menzel, C.M. Schneider, R. Waser, and R. Dittmann, Nat. Commun. **7**, 12398 (2016).

[82] S. Nemšák, E. Strelcov, T. Duchoň, H. Guo, J. Hackl, A. Yulaev, I. Vlassiouk, D.N. Mueller, C.M. Schneider, and A. Kolmakov, J. Am. Chem. Soc. **139**, 18138 (2017).

[83] H. Guo, E. Strelcov, A. Yulaev, J. Wang, N. Appathurai, S. Urquhart, J. Vinson, S. Sahu, M. Zwolak, and A. Kolmakov, Nano Lett. **17**, 1034 (2017).

[84] S. Nemšák, E. Strelcov, H. Guo, B.D. Hoskins, T. Duchoň, D.N. Mueller, A. Yulaev, I. Vlassiouk, A. Tselev, C.M. Schneider, and A. Kolmakov, Top. Catal. **61**, 2195 (2018).

[85] O. Karslıoğlu, S. Nemšák, I. Zegkinoglou, A. Shavorskiy, M. Hartl, F. Salmassi, E.M. Gullikson, M.L. Ng, C. Rameshan, B. Rude, D. Bianculli, A.A. Cordones, S. Axnanda, E.J. Crumlin, P.N. Ross, C.M. Schneider, Z. Hussain, Z. Liu, C.S. Fadley, and H. Bluhm, Faraday Discuss. **180**, 35 (2015).

[86] S. Nemšák, A. Shavorskiy, O. Karslioglu, I. Zegkinoglou, A. Rattanachata, C.S. Conlon, A. Keqi, P.K.

Greene, E.C. Burks, F. Salmassi, E.M. Gullikson, S.-H. Yang, K. Liu, H. Bluhm, and C.S. Fadley, Nat. Commun. **5**, 5441 (2014).

[87] A.X. Gray, S. Nemšák, and C.S. Fadley, Synchrotron Radiat. News **31**, 42 (2018).

[88] S. Nemšák, A.X. Gray, and C.S. Fadley, in *Spectrosc. Complex Oxide Interfaces Photoemiss. Relat. Spectrosc.*, edited by C. Cancellieri and V.N. Strocov (Springer International Publishing, Cham, 2018), pp. 153–179.

[89] M. Setvín, M. Wagner, M. Schmid, G.S. Parkinson, and U. Diebold, Chem. Soc. Rev. **46**, 1772 (2017).

[90] G. Binnig, C.F. Quate, and C. Gerber, Phys. Rev. Lett. **56**, 930 (1986).

[91] L. Österlund, P.B. Rasmussen, P. Thostrup, E. Lægsgaard, I. Stensgaard, and F. Besenbacher, Phys. Rev. Lett. **86**, 460 (2001).

[92] J. Guo, K. Bian, Z. Lin, and Y. Jiang, J. Chem. Phys. **145**, 160901 (2016).

[93] Y. Liang, J.H.K. Pfisterer, D. McLaughlin, C. Csoklich, L. Seidl, A.S. Bandarenka, and O. Schneider, Small Methods **3**, 1800387 (2019).

[94] C.L. Bentley, J. Edmondson, G.N. Meloni, D. Perry, V. Shkirskiy, and P.R. Unwin, Anal. Chem. **91**, 84 (2019).

[95] X. Li, H.-Y. Wang, H. Yang, W. Cai, S. Liu, and B. Liu, Small Methods **2**, 1700395 (2018).

[96] Y.-S. Liu, X. Feng, P.-A. Glans, and J. Guo, Sol. Energy Mater. Sol. Cells **208**, 110432 (2020).

[97] F. de Groot, Chem. Rev. **101**, 1779 (2001).

[98] M. Fleischmann and B.W. Mao, J. Electroanal. Chem. Interfacial Electrochem. **229**, 125 (1987).

[99] F. Reikowski, F. Maroun, I. Pacheco, T. Wiegmann, P. Allongue, J. Stettner, and O.M. Magnussen, ACS Catal. **9**, 3811 (2019).

[100] J. Drnec, M. Ruge, F. Reikowski, B. Rahn, F. Carlà, R. Felici, J. Stettner, O.M. Magnussen, and D.A. Harrington, Electrochim. Acta **224**, 220 (2017).

[101] O.M. Magnussen, K. Krug, A.H. Ayyad, and J. Stettner, Electrochim. Acta **53**, 3449 (2008).

[102] R.R. Rao, M.J. Kolb, L. Giordano, A.F. Pedersen, Y. Katayama, J. Hwang, A. Mehta, H. You, J.R. Lunger, H. Zhou, N.B. Halck, T. Vegge, I. Chorkendorff, I.E.L. Stephens, and Y. Shao-Horn, Nat. Catal. **3**, 516 (2020).

[103] R.R. Rao, M.J. Kolb, N.B. Halck, A.F. Pedersen, A. Mehta, H. You, K.A. Stoerzinger, Z. Feng, H.A.

Hansen, H. Zhou, L. Giordano, J. Rossmeisl, T. Vegge, I. Chorkendorff, I.E.L. Stephens, and Y. Shao-Horn, Energy Environ. Sci. **10**, 2626 (2017).

[104] T. Schmitt, F.M.F. de Groot, and J.-E. Rubensson, J. Synchrotron Radiat. **21**, 1065 (2014).

[105] S. Basak, V. Migunov, A.H. Tavabi, C. George, Q. Lee, P. Rosi, V. Arszelewska, S. Ganapathy, A. Vijay, F. Ooms, R. Schierholz, H. Tempel, H. Kungl, J. Mayer, R.E. Dunin-Borkowski, R.-A. Eichel, M. Wagemaker, and E.M. Kelder, ACS Appl. Energy Mater. **3**, 5101 (2020).

[106] G. Lole, V. Roddatis, U. Ross, M. Risch, T. Meyer, L. Rump, J. Geppert, G. Wartner, P. Blöchl, and C. Jooss, Commun. Mater. **1**, 68 (2020).

[107] Roddatis, Lole, and Jooss, Catalysts **9**, 751 (2019).

[108] R. Sinclair, MRS Bull. **38**, 1065 (2013).

[109] T.W. Hansen, *Controlled Atmosphere Transmission Electron Microscopy* (Springer International Publishing, Cham, 2016).

[110] M.A. Ali, A. Hassan, G.C. Sedenho, R. V. Gonçalves, D.R. Cardoso, and F.N. Crespilho, J. Phys. Chem. C **123**, 16058 (2019).

[111] M. Sathiya, J.-B. Leriche, E. Salager, D. Gourier, J.-M. Tarascon, and H. Vezin, Nat. Commun. **6**, 6276 (2015).

[112] Y.J. Tong, Curr. Opin. Electrochem. **4**, 60 (2017).

[113] N.L. Yamada, T. Hosobata, F. Nemoto, K. Hori, M. Hino, J. Izumi, K. Suzuki, M. Hirayama, R. Kanno, and Y. Yamagata, J. Appl. Crystallogr. **53**, 1462 (2020).

[114] R. Mendelsohn, G. Mao, and C.R. Flach, Biochim. Biophys. Acta - Biomembr. **1798**, 788 (2010).

[115] F. Faisal, C. Stumm, M. Bertram, F. Waidhas, Y. Lykhach, S. Cherevko, F. Xiang, M. Ammon, M. Vorokhta, B. Šmíd, T. Skála, N. Tsud, A. Neitzel, K. Beranová, K.C. Prince, S. Geiger, O. Kasian, T. Wähler, R. Schuster, M.A. Schneider, V. Matolín, K.J.J. Mayrhofer, O. Brummel, and J. Libuda, Nat. Mater. **17**, 592 (2018).

[116] M. Osawa, M. Tsushima, H. Mogami, G. Samjeské, and A. Yamakata, J. Phys. Chem. C **112**, 4248 (2008).

[117] Z.-Q. Tian and B. Ren, Annu. Rev. Phys. Chem. **55**, 197 (2004).

[118] D.B. O'Neill, D. Prezgot, A. Ianoul, C. Otto, G. Mul, and A. Huijser, ACS Appl. Mater. Interfaces **12**, 1905 (2020).

[119] X. Wang, Z. Wang, F.P. García de Arquer, C.-T. Dinh, A. Ozden, Y.C. Li, D.-H. Nam, J. Li, Y.-S. Liu, J. Wicks, Z. Chen, M. Chi, B. Chen, Y. Wang, J. Tam, J.Y. Howe, A. Proppe, P. Todorović, F. Li, T.-T. Zhuang, C.M. Gabardo, A.R. Kirmani, C. McCallum, S.-F. Hung, Y. Lum, M. Luo, Y. Min, A. Xu, C.P. O'Brien, B. Stephen, B. Sun, A.H. Ip, L.J. Richter, S.O. Kelley, D. Sinton, and E.H. Sargent, Nat. Energy **5**, 478 (2020).

[120] T. Homma, M. Kunimoto, M. Sasaki, T. Hanai, and M. Yanagisawa, J. Appl. Electrochem. **48**, 561 (2018).

[121] J. Stöhr, *NEXAFS Spectroscopy* (Springer Berlin Heidelberg, Berlin, Heidelberg, 1992).

[122] R.W. Johnston and D.H. Tomboulian, Phys. Rev. **94**, 1585 (1954).

[123] B.L. Henke, E.M. Gullikson, and J.C. Davis, At. Data Nucl. Data Tables **54**, 181 (1993).

[124] D. Friebel, D.J. Miller, H. Ogasawara, T. Anniyev, U. Bergmann, J. Bargar, and A. Nilsson, (2009).

[125] X. Zhang, G. Smolentsev, J. Guo, K. Attenkofer, C. Kurtz, G. Jennings, J. V. Lockard, A.B. Stickrath, and L.X. Chen, J. Phys. Chem. Lett. **2**, 628 (2011).

[126] C. Schmitz-Antoniak, D. Schmitz, A. Warland, M. Darbandi, S. Haldar, S. Bhandary, B. Sanyal, O. Eriksson, and H. Wende, Ann. Phys. **530**, 1700363 (2018).

[127] Q. Lu, Y. Chen, H. Bluhm, and B. Yildiz, J. Phys. Chem. C **120**, 24148 (2016).

[128] S.-C. Lin, C.-C. Chang, S.-Y. Chiu, H.-T. Pai, T.-Y. Liao, C.-S. Hsu, W.-H. Chiang, M.-K. Tsai, and H.M. Chen, Nat. Commun. **11**, 3525 (2020).

[129] L. Braglia, M. Fracchia, P. Ghigna, A. Minguzzi, D. Meroni, R. Edla, M. Vandichel, E. Ahlberg, G. Cerrato, and P. Torelli, J. Phys. Chem. C **124**, 14202 (2020).

[130] D. Chen, Z. Guan, D. Zhang, L. Trotochaud, E. Crumlin, S. Nemsak, H. Bluhm, H.L. Tuller, and W.C. Chueh, Nat. Catal. **3**, 116 (2020).

[131] D.N. Mueller, M.L. Machala, H. Bluhm, and W.C. Chueh, Nat. Commun. **6**, 6097 (2015).

[132] D. Drevon, M. Görlin, P. Chernev, L. Xi, H. Dau, and K.M. Lange, Sci. Rep. **9**, 1 (2019).

[133] C. Schwanke, L. Xi, and K.M. Lange, J. Synchrotron Radiat. **23**, 1390 (2016).

[134] E. Fabbri, M. Nachtegaal, T. Binninger, X. Cheng, B.-J. Kim, J. Durst, F. Bozza, T. Graule, R. Schäublin, L. Wiles, M. Pertoso, N. Danilovic, K.E. Ayers, and T.J. Schmidt, Nat. Mater. **16**, 925 (2017).

[135] L.J.A. Macedo, A. Hassan, G.C. Sedenho, and F.N. Crespilho, Nat. Commun. **11**, (2020).

[136] M. Al Samarai, A.W. Hahn, A. Beheshti Askari, Y.-T. Cui, K. Yamazoe, J. Miyawaki, Y. Harada, O. Rüdiger, and S. DeBeer, ACS Appl. Mater. Interfaces acsami.9b06752 (2019).

[137] C. Liu, J. Qian, Y. Ye, H. Zhou, C. Sun, C. Sheehan, Z. Zhang, G. Wan, Y. Liu, J. Guo, S. Li, H. Shin, S. Hwang, T.B. Gunnoe, W.A.G. Iii, and S. Zhang, Nat. Catal. (n.d.).

[138] Y. Gorlin, M.U.M. Patel, A. Freiberg, Q. He, M. Piana, M. Tromp, and H.A. Gasteiger, J. Electrochem. Soc. **163**, A930 (2016).

[139] P. Zimmermann, S. Peredkov, P.M. Abdala, S. DeBeer, M. Tromp, C. Müller, and J.A. van Bokhoven, Coord. Chem. Rev. **423**, 213466 (2020).

[140] W.S.M. Werner, Physics (College. Park. Md). **12**, 93 (2019).

[141] N. Ottosson, M. Faubel, S.E. Bradforth, P. Jungwirth, and B. Winter, J. Electron Spectros. Relat. Phenomena **177**, 60 (2010).

[142] S. Tanuma, C.J. Powell, and D.R. Penn, Surf. Interface Anal. **21**, 165 (1994).

[143] J.E. Kleibeuker, Z. Zhong, H. Nishikawa, J. Gabel, A. Müller, F. Pfaff, M. Sing, K. Held, R. Claessen, G. Koster, and G. Rijnders, Phys. Rev. Lett. **113**, 237402 (2014).

[144] Y. Hermans, S. Murcia-López, A. Klein, R. van de Krol, T. Andreu, J.R. Morante, T. Toupance, and W. Jaegermann, Phys. Chem. Chem. Phys. **21**, 5086 (2019).

[145] B. Arndt, F. Borgatti, F. Offi, M. Phillips, P. Parreira, T. Meiners, S. Menzel, K. Skaja, G. Panaccione, D.A. MacLaren, R. Waser, and R. Dittmann, Adv. Funct. Mater. **27**, 1702282 (2017).

[146] C. Guhl, P. Kehne, Q. Ma, F. Tietz, L. Alff, P. Komissinskiy, W. Jaegermann, and R. Hausbrand, Rev. Sci. Instrum. **89**, 073104 (2018).

[147] R. Endo, T. Ohnishi, K. Takada, and T. Masuda, J. Phys. Chem. Lett. **11**, 6649 (2020).

[148] H. Siegbahn, S. Svensson, and M. Lundholm, J. Electron Spectros. Relat. Phenomena **24**, 205 (1981).

[149] H. Siegbahn and K. Siegbahn, J. Electron Spectros. Relat. Phenomena **2**, 319 (1973).

[150] D.F. Ogletree, H. Bluhm, G. Lebedev, C.S. Fadley, Z. Hussain, and M. Salmeron, Rev. Sci. Instrum. **73**, 3872 (2002).

[151] H. Bluhm, M. Hävecker, A. Knop-Gericke, E. Kleimenov, R. Schlögl, D. Teschner, V.I. Bukhtiyarov, D.F. Ogletree, and M. Salmeron, J. Phys. Chem. B **108**, 14340 (2004).

[152] H. Bluhm, Top. Catal. **59**, 403 (2016).

[153] J.T. Newberg, J. Åhlund, C. Arble, C. Goodwin, Y. Khalifa, and A. Broderick, Rev. Sci. Instrum. **86**, 085113 (2015).

[154] C. Arble, M. Jia, and J.T. Newberg, Surf. Sci. Rep. **73**, 37 (2018).

[155] J.S. Elias, K.A. Stoerzinger, W.T. Hong, M. Risch, L. Giordano, A.N. Mansour, and Y. Shao-Horn, ACS Catal. **7**, 6843 (2017).

[156] B. Eren and A.R. Head, J. Phys. Chem. C **124**, 3557 (2020).

[157] F. Tao, S. Dag, L.-W. Wang, Z. Liu, D.R. Butcher, H. Bluhm, M. Salmeron, and G.A. Somorjai, Science **327**, 850 (2010).

[158] C.J. (Kee. (Kees-J. Weststrate, D. Sharma, D. Garcia Rodriguez, M.A. Gleeson, H.O.A. Fredriksson, and J.W. (Hans) Niemantsverdriet, Nat. Commun. **11**, 750 (2020).

[159] Z. Guan, D. Chen, and W.C. Chueh, Phys. Chem. Chem. Phys. **19**, 23414 (2017).

[160] C. Balaji Gopal, M. García-Melchor, S.C. Lee, Y. Shi, A. Shavorskiy, M. Monti, Z. Guan, R. Sinclair, H. Bluhm, A. Vojvodic, and W.C. Chueh, Nat. Commun. **8**, 15360 (2017).

[161] H.S. Casalongue, S. Kaya, V. Viswanathan, D.J. Miller, D. Friebel, H.A. Hansen, J.K. Nørskov, A. Nilsson, and H. Ogasawara, Nat. Commun. **4**, 2817 (2013).

[162] S. Axnanda, M. Scheele, E. Crumlin, B. Mao, R. Chang, S. Rani, M. Faiz, S. Wang, A.P. Alivisatos, and Z. Liu, Nano Lett. **13**, 6176 (2013).

[163] M. Andrä, H. Bluhm, R. Dittmann, C.M. Schneider, R. Waser, D.N. Mueller, and F. Gunkel, Phys. Rev. Mater. **3**, 044604 (2019).

[164] M. Rose, B. Šmíd, M. Vorokhta, I. Slipukhina, M. Andrä, H. Bluhm, T. Duchoň, M. Ležaić, S.A. Chambers, R. Dittmann, D.N. Mueller, and F. Gunkel, Adv. Mater. **2004132**, 2004132 (2020).

[165] J. Cai, Q. Dong, Y. Han, B.-H. Mao, H. Zhang, P.G. Karlsson, J. Åhlund, R.-Z. Tai, Y. Yu, and Z. Liu, Nucl. Sci. Tech. **30**, 81 (2019).

[166] J.O. Bockris and B.D. Cahan, J. Chem. Phys. **50**, 1307 (1969).

[167] D. Weingarth, A. Foelske-Schmitz, A. Wokaun, and R. Kötz, Electrochem. Commun. **13**, 619 (2011).

[168] S.G. Booth, A.M. Tripathi, I. Strashnov, R.A.W. Dryfe, and A.S. Walton, J. Phys. Condens. Matter **29**, 454001 (2017).

[169] M.F. Lichterman, S. Hu, M.H. Richter, E.J. Crumlin, S. Axnanda, M. Favaro, W. Drisdell, Z. Hussain, T. Mayer, B.S. Brunschwig, N.S. Lewis, Z. Liu, and H.-J. Lewerenz, Energy Environ. Sci. **8**, 2409 (2015).

[170] A. Shavorskiy, X. Ye, O. Karslıoğlu, A.D. Poletayev, M. Hartl, I. Zegkinoglou, L. Trotochaud, S. Nemšák, C.M. Schneider, E.J. Crumlin, S. Axnanda, Z. Liu, P.N. Ross, W. Chueh, and H. Bluhm, J. Phys. Chem. Lett. **8**, 5579 (2017).

[171] M. Favaro, C. Valero-Vidal, J. Eichhorn, F.M. Toma, P.N. Ross, J. Yano, Z. Liu, and E.J. Crumlin, J. Mater. Chem. A **5**, 11634 (2017).

[172] K.A. Stoerzinger, M. Favaro, P.N. Ross, J. Yano, Z. Liu, Z. Hussain, and E.J. Crumlin, J. Phys. Chem. B **122**, 864 (2018).

[173] H. Ali-Löytty, M.W. Louie, M.R. Singh, L. Li, H.G. Sanchez Casalongue, H. Ogasawara, E.J. Crumlin, Z. Liu, A.T. Bell, A. Nilsson, and D. Friebel, J. Phys. Chem. C **120**, 2247 (2016).

[174] M. Favaro, J. Yang, S. Nappini, E. Magnano, F.M. Toma, E.J. Crumlin, J. Yano, and I.D. Sharp, J. Am. Chem. Soc. **139**, 8960 (2017).

[175] M. Favaro, W.S. Drisdell, M.A. Marcus, J.M. Gregoire, E.J. Crumlin, J.A. Haber, and J. Yano, ACS Catal. **7**, 1248 (2017).

[176] K.A. Stoerzinger, X. Renshaw Wang, J. Hwang, R.R. Rao, W.T. Hong, C.M. Rouleau, D. Lee, Y. Yu, E.J. Crumlin, and Y. Shao-Horn, Top. Catal. **61**, 2161 (2018).

[177] R.S. Weatherup, C.H. Wu, C. Escudero, V. Pérez-Dieste, and M.B. Salmeron, J. Phys. Chem. B **122**, 737 (2018).

[178] Z. Nagy and H. You, J. Electroanal. Chem. **381**, 275 (1995).

[179] J.J. Velasco-Velez, C.H. Wu, B.Y. Wang, Y. Sun, Y. Zhang, J.-H. Guo, and M. Salmeron, J. Phys. Chem. C **118**, 25456 (2014).

[180] Y.-H. Lu, C. Morales, X. Zhao, M.A. van Spronsen, A. Baskin, D. Prendergast, P. Yang, H.A. Bechtel, E.S. Barnard, D.F. Ogletree, V. Altoe, L. Soriano, A.M. Schwartzberg, and M. Salmeron, Nano Lett. **20**, 6364 (2020).

[181] A. Kolmakov, D.A. Dikin, L.J. Cote, J. Huang, M.K. Abyaneh, M. Amati, L. Gregoratti, S. Günther, and M. Kiskinova, Nat. Nanotechnol. **6**, 651 (2011).

[182] J. Kraus, R. Reichelt, S. Günther, L. Gregoratti, M. Amati, M. Kiskinova, A. Yulaev, I. Vlassiouk, and A. Kolmakov, Nanoscale **6**, 14394 (2014).

[183] R.S. Weatherup, Top. Catal. **61**, 2085 (2018).

[184] Y.T. Law, S. Zafeiratos, S.G. Neophytides, A. Orfanidi, D. Costa, T. Dintzer, R. Arrigo, A. Knop-

Gericke, R. Schlögl, and E.R. Savinova, Chem. Sci. **6**, 5635 (2015).

[185] L.J. Falling, R. V. Mom, L.E. Sandoval Diaz, S. Nakhaie, E. Stotz, D. Ivanov, M. Hävecker, T. Lunkenbein, A. Knop-Gericke, R. Schlögl, and J.-J. Velasco-Vélez, ACS Appl. Mater. Interfaces **12**, 37680 (2020).

[186] C. Baeumer, R. Valenta, C. Schmitz, A. Locatelli, T.O. Menteş, S.P. Rogers, A. Sala, N. Raab, S. Nemsak, M. Shim, C.M. Schneider, S. Menzel, R. Waser, and R. Dittmann, ACS Nano **11**, 6921 (2017).

[187] L.A. Baker, J. Am. Chem. Soc. (2018).

[188] E. Strelcov, C. Arble, H. Guo, B.D. Hoskins, A. Yulaev, I. V. Vlassiouk, N.B. Zhitenev, A. Tselev, and A. Kolmakov, Nano Lett. **20**, 1336 (2020).

[189] A. Labuda, F. Hausen, N.N. Gosvami, P.H. Grütter, R.B. Lennox, and R. Bennewitz, Langmuir **27**, 2561 (2011).

[190] J. Walter, G. Yu, B. Yu, A. Grutter, B. Kirby, J. Borchers, Z. Zhang, H. Zhou, T. Birol, M. Greven, and C. Leighton, Phys. Rev. Mater. **1**, 071403 (2017).

[191] G. Eres, C.M. Rouleau, Q. Lu, Z. Zhang, E. Benda, H.N. Lee, J.Z. Tischler, and D.D. Fong, Rev. Sci. Instrum. **90**, 093902 (2019).

[192] H. Kersell, P. Chen, H. Martins, Q. Lu, F. Brausse, B.-H. Liu, M. Blum, S. Roy, B. Rude, A. Kilcoyne, H. Bluhm, and S. Nemšák, 1 (2021).

[193] B. Venderbosch, J.-P.H. Oudsen, L.A. Wolzak, D.J. Martin, T.J. Korstanje, and M. Tromp, ACS Catal. **9**, 1197 (2019).

[194] J.P.H. Oudsen, B. Venderbosch, D.J. Martin, T.J. Korstanje, J.N.H. Reek, and M. Tromp, Phys. Chem. Chem. Phys. **21**, 14638 (2019).

[195] V. Pfeifer, T.E. Jones, J.J. Velasco Vélez, R. Arrigo, S. Piccinin, M. Hävecker, A. Knop-Gericke, and R. Schlögl, Chem. Sci. **8**, 2143 (2017).

[196] F.D. Speck, P.G. Santori, F. Jaouen, and S. Cherevko, J. Phys. Chem. C **123**, 25267 (2019).

[197] S. Geiger, O. Kasian, M. Ledendecker, E. Pizzutilo, A.M. Mingers, W.T. Fu, O. Diaz-Morales, Z. Li, T. Oellers, L. Fruchter, A. Ludwig, K.J.J. Mayrhofer, M.T.M. Koper, and S. Cherevko, Nat. Catal. **1**, 508 (2018).

[198] J. Liu, E. Jia, K.A. Stoerzinger, L. Wang, Y. Wang, Z. Yang, D. Shen, M.H. Engelhard, M.E. Bowden, Z. Zhu, S.A. Chambers, and Y. Du, J. Phys. Chem. C **124**, 15386 (2020).

[199] M. Risch, A. Grimaud, K.J. May, K.A. Stoerzinger, T.J. Chen, A.N. Mansour, and Y. Shao-Horn, J. Phys. Chem. C **117**, 8628 (2013).

[200] D. Antipin and M. Risch, J. Phys. Energy **2**, 032003 (2020).

[201] M.L. Weber and F. Gunkel, J. Phys. Energy **1**, 031001 (2019).

[202] M. Risch, Catalysts **7**, 154 (2017).

[203] K.A. Stoerzinger, M. Risch, J. Suntivich, W.M. Lü, J. Zhou, M.D. Biegalski, H.M. Christen, Ariando, T. Venkatesan, and Y. Shao-Horn, Energy Environ. Sci. **6**, 1582 (2013).

[204] W. Smekal, W.S.M. Werner, and C.J. Powell, Surf. Interface Anal. **37**, 1059 (2005).

[205] J.T. Mefford, X. Rong, A.M. Abakumov, W.G. Hardin, S. Dai, A.M. Kolpak, K.P. Johnston, and K.J. Stevenson, Nat. Commun. **7**, 11053 (2016).

[206] K.J. May, C.E. Carlton, K.A. Stoerzinger, M. Risch, J. Suntivich, Y.-L. Lee, A. Grimaud, and Y. Shao-Horn, J. Phys. Chem. Lett. **3**, 3264 (2012).

[207]MAX IV HIPPIE Beamline (2018).

[208] M.C. Biesinger, B.P. Payne, A.P. Grosvenor, L.W.M. Lau, A.R. Gerson, and R.S.C. Smart, Appl. Surf. Sci. **257**, 2717 (2011).

[209] Y. Zhang, Y. Khalifa, E.J. Maginn, and J.T. Newberg, J. Phys. Chem. C **122**, 27392 (2018).

[210] M. Pijolat and G. Hollinger, Surf. Sci. **105**, 114 (1981).

[211] R. Claessen, M. Sing, M. Paul, G. Berner, A. Wetscherek, A. Müller, and W. Drube, New J. Phys. **11**, 125007 (2009).

[212] R.W. Paynter, J. Electron Spectros. Relat. Phenomena **169**, 1 (2009).

[213] C. Baeumer, J. Li, Q. Lu, A.Y.-L. Liang, L. Jin, H.P. Martins, T. Duchoň, M. Glöß, S.M. Gericke, M.A. Wohlgemuth, M. Giesen, E.E. Penn, R. Dittmann, F. Gunkel, R. Waser, M. Bajdich, S. Nemšák, J.T. Mefford, and W.C. Chueh, Nat. Mater. (2021).

[214] H.P. Martins, S.A. Khan, G. Conti, A.A. Greer, A.Y. Saw, G.K. Palsson, M. Huijben, K. Kobayashi, S. Ueda, C.M. Schneider, I.M. Vishik, J. Minár, A.X. Gray, C.S. Fadley, and S. Nemšák, (2020).

[215] C. Baeumer, S.P. Rogers, R. Xu, L.W. Martin, and M. Shim, Nano Lett. **13**, 1693 (2013).

[216] Y. Ning, Y. Li, C. Wang, R. Li, F. Zhang, S. Zhang, Z. Wang, F. Yang, N. Zong, Q. Peng, Z. Xu, X. Wang, R. Li, M. Breitschaft, S. Hagen, O. Schaff, Q. Fu, and X. Bao, Rev. Sci. Instrum. **91**, 113704 (2020).

[217] K. Roy, J. Raabe, P. Schifferle, S. Finizio, A. Kleibert, J.A. van Bokhoven, and L. Artiglia, J. Synchrotron Radiat. **26**, 785 (2019).

[218] F. Roth, S. Neppl, A. Shavorskiy, T. Arion, J. Mahl, H.O. Seo, H. Bluhm, Z. Hussain, O. Gessner, and W. Eberhardt, Phys. Rev. B **99**, 020303 (2019).

[219] R. Brüninghoff, K. Wenderich, J.P. Korterik, B.T. Mei, G. Mul, and A. Huijser, J. Phys. Chem. C **123**, 26653 (2019).

[220] A. Huijser, Q. Pan, D. van Duinen, M.G. Laursen, A. El Nahhas, P. Chabera, L. Freitag, L. González, Q. Kong, X. Zhang, K. Haldrup, W.R. Browne, G. Smolentsev, and J. Uhlig, J. Phys. Chem. A **122**, 6396 (2018).

[221] S. Kumar, C.E. Graves, J.P. Strachan, A.L.D. Kilcoyne, T. Tyliszczak, Y. Nishi, and R.S. Williams, J. Appl. Phys. **118**, 034502 (2015).